\documentclass[10pt]{iopart}

\usepackage{iopams}
\usepackage[T1]{fontenc}
\usepackage[utf8]{inputenc}
\usepackage{textcomp}
\usepackage{float}
\usepackage{graphicx}
\usepackage{tabularx}
\usepackage{subfigure}
\usepackage{grffile}
\usepackage{caption}
\usepackage[dvipsnames]{xcolor}
\usepackage{ulem}
\usepackage{soul}
\usepackage{enumerate}
\usepackage{hyperref}
\usepackage{url}
\usepackage{cite}
\usepackage{comment}
\usepackage{todonotes}
\usepackage{multirow}
\usepackage{booktabs}

\graphicspath{{./Figures/}}

\setulcolor{blue}
\setstcolor{red}

\soulregister\cite7
\soulregister\ref7
\soulregister\eref7
\soulregister\Eref7
\soulregister\fref7
\soulregister\Fref7
\soulregister\sref7
\soulregister\Sref7
\soulregister\tref7
\soulregister\Tref7
\soulregister\pageref7

\begin{document}

\title{A computational study of accelerating, steady and fading negative streamers in ambient air}

\author{Baohong Guo$^{1}$, Xiaoran Li$^{1,2}$, Ute Ebert$^{1,3}$, Jannis Teunissen$^{1,*}$}

\address{$^1$ Centrum Wiskunde \& Informatica (CWI), Amsterdam, The Netherlands}
\address{$^2$ State Key Laboratory of Electrical Insulation and Power Equipment, School of Electrical Engineering, Xi’an Jiaotong University, Xi’an, 710049, People's Republic of China}
\address{$^3$ Department of Applied Physics, Eindhoven University of Technology, Eindhoven, The Netherlands}

\ead{jannis.teunissen@cwi.nl}

\vspace{10pt}

\begin{indented}
\item[]
\today
\end{indented}

\begin{abstract}
  We study negative streamers in ambient air using a 2D axisymmetric fluid model.
  Depending on the background electric field, we observe accelerating, steady and fading negative streamers.
  Fading occurs in low background fields, when negative streamers lose their field enhancement and when their velocities become comparable to their maximal electron drift velocities.
  Our focus is on the steady propagation mode, during which streamer properties like radius and velocity hardly change.
  However, this mode is unstable, in the sense that a small change in conditions leads to acceleration or deceleration.
  We observe steady negative streamers in background fields ranging from 9.19\,kV/cm to 15.75\,kV/cm, indicating that there is no unique steady propagation field (or stability field).
  Another finding is that steady negative streamers are able to keep propagating over tens of centimeters, with only a finite conductive length behind their heads, similar to steady positive streamers.
  Approximately linear relationships are observed between the optical diameter and properties like the streamer velocity and the streamer head potential.
  From these linear relations, we obtain rough lower bounds of about 0.27\,mm to 0.35\,mm for the minimal optical diameter of steady negative streamers.
  The lowest background field in which a steady negative streamer could be obtained is 9.19\,kV/cm.
  In contrast, steady positive streamers have recently been obtained in a background field as low as 4.05\,kV/cm.
  We find that the properties of steady negative and positive streamers differ significantly.
  For example, for steady negative streamers the ratio between streamer velocity and maximal electron drift velocity ranges from about 2 to 4.5, whereas for steady positive streamers this ratio ranges from about 0.05 to 0.26.
  % The screening of the internal electric field is much weaker for steady negative streamers.
\end{abstract}

% \vspace{2pc}
% \noindent{\it Keywords}: negative streamer, negative streamer fading, streamer stagnation, steady streamer propagation, stability field
% \submitto{\PSST on April 6, 2022}
%\maketitle
\ioptwocol

\section{Introduction}\label{sec:introduction}

Streamer discharges are fast-moving ionization fronts with self-organized field enhancement at their tips~\cite{nijdam2020a}.
This field enhancement allows them to propagate into regions where the background electric field is below breakdown~\cite{huiskamp2020}.
Streamers are the precursors of sparks and lightning leaders~\cite{gallimberti1979a, lalande2002}, and they exist in nature as so-called sprites~\cite{pasko2007, ebert2010}. Streamers are widely observed in cold atmospheric plasma applications~\cite{kim2004, keidar2013, starikovskaia2014}, as well as in high-voltage technology~\cite{Raizer1991, veldhuizen2000, seeger2018}.

Streamers can have positive or negative polarity.
Positive streamers in air have been studied and modelled more frequently than negative ones because they initiate more easily~\cite{briels2008, starikovskiy2020}.
However, negative streamers also widely exist in nature and industry~\cite{reess1995, taylor2008, arevalo2011a, kochkin2014, malagon-romero2019, tilles2019, koile2020a, scholten2021}.
In experimental~\cite{yi2002, wang2007, briels2008, huiskamp2017} and computational~\cite{babaeva1997, liu2004, luque2008, naidis2009, babaeva2016a, starikovskiy2020} comparisons between positive and negative streamers in the same background field, positive streamers in atmospheric-pressure air were typically faster and thicker than negative ones, with higher field enhancement at their heads and higher plasma densities in their channels.
In higher background fields, these differences were smaller.

In this paper, we study the dynamics of negative streamer propagation in air with simulations.
We focus on three main topics.
First, we study negative streamer fading dynamics in weak background fields.
Second, we look into steady streamer propagation, i.e., propagation with a constant velocity during which streamer properties hardly change.
We study in particular how the properties of steady negative streamers vary with the background field.
Third, we consider the relationships between steady negative streamer properties like diameter, velocity and maximal electric field, and we compare these properties to those of steady positive streamers.

There is a difference between negative streamer fading and positive streamer stagnation.
A fading negative streamer loses its field enhancement, but electrons near the streamer head continue their drift motion.
In contrast, a stagnating positive streamer will eventually come to an approximate halt due to the relative immobility of positive ions~\cite{niknezhad2021b,li2022}, and its field enhancement tends to increase before stagnation.

There are relatively few experimental studies in which the propagation of negative streamers in air was captured.
Most relevant for the present paper is the work of Briels \textit{et al}~\cite{briels2008}, in which positive and negative streamers were studied at various voltages in a needle-to-plane gap of 4\,cm.
Negative streamers could form at voltages above 30\,kV, and they could cross the gap at voltages above 56\,kV.
Negative streamers in air were also captured in~\cite{kochkin2014}, on microsecond time scales in a meter-sized gap, and in~\cite{winands2008}, in a wire-plate geometry.

Several authors have numerically studied the propagation of negative streamers.
In~\cite{babaeva1997}, negative streamer deceleration was simulated in atmospheric air in weak uniform electric fields.
In the numerical study of~\cite{luque2008}, a negative discharge in air at standard temperature and pressure was found to extinguish after less than 2\,mm propagation in a low background field.
However, the authors noted that a negative discharge of only 2\,mm cannot be considered a fully developed streamer.

Recently, Starikovskiy \textit{et al}~\cite{starikovskiy2021} simulated decelerating streamers in 5\,cm and 14\,cm gaps for both voltage polarities.
Regardless of the gap geometry, the authors found that a negative streamer in air decelerated with a decrease in the maximal electric field and an increase in the radius, and it eventually transformed to a discharge mode dominated by electron drift.
Furthermore, they showed that the deceleration dynamics of negative and positive streamers were completely different.

From a numerical modeling point of view, negative streamer fading is easier to simulate than positive streamer stagnation.
As pointed out in~\cite{pancheshnyi2004a, starikovskiy2021, francisco2021e}, the electric field can diverge when positive streamer stagnation is simulated using a fluid model with the local field approximation.
Such problems can be avoided by using an extended fluid model~\cite{niknezhad2021b} or by modifying the impact ionization source term~\cite{li2022}.

To understand streamer propagation, the phenomenological concept of a `stability field' has frequently been used~\cite{phelps1971a, gallimberti1979a, allen1995, nijdam2020a}.
The stability field was defined as the background electric field required for steady streamer propagation~\cite{babaeva1997, qin2014, francisco2021e, li2022}.
This field is here referred to as `the steady propagation field'.
More commonly, the stability field has been defined as the minimal electric field required for a streamer to cross a gap~\cite{phelps1971a, phelps1976, griffiths1976a, gallimberti1979a}.
As pointed out in~\cite{phelps1976, griffiths1976a, allen1991, niemeyer1995, allen1999, qin2014, li2022}, both the steady propagation field and the stability field are not unique.
They depend not only on factors such as the gas, humidity, and applied voltage polarity, but also on streamer properties like the velocity and radius.
We will discuss the steady propagation field and the stability field in more detail in \sref{sec:stab-fields-steady}.

Several observations have been made about the velocity--diameter relation of negative streamers in air.
Experimentally, an empirical fit $v=0.5d^2\, \mathrm{mm}^{-1}\,\mathrm{s}^{-1}$ for both positive and negative streamers in ambient air at 1 bar was obtained in~\cite{briels2008}.
In contrast, a linear relationship between velocity and diameter was predicted in the analytical and numerical study of~\cite{naidis2009}, if the maximal electric field at the streamer head was assumed constant, and if the relative increase of electron density at the streamer tip was fixed.
However, deviations from linearity are to be expected, since the maximal electric field depends on factors such as the streamer radius and the applied voltage, and since the relative electron density increase can also vary.

The present study was performed in parallel with the one on steady positive streamers described in \cite{li2022}.
In both papers steady streamers with different properties were obtained, but using different methods.
In~\cite{li2022}, the applied voltage was changed in time to obtain certain constant streamer velocities.
Here we instead use a DC applied voltage (constant in time).
A voltage corresponding to steady propagation is found by performing simulations with different applied voltages.
For given electrode and initial conditions we only find one such voltage, so we vary the electrode geometry to obtain different steady negative streamers.

The paper is organized as follows.
In \sref{sec:simulation model}, we describe the 2D axisymmetric fluid model, along with the chemical reactions and simulation conditions.
The simulation results are presented in \sref{sec:simulation results}.
Accelerating, steady and fading negative streamers are investigated in different background fields.
Then we discuss stability fields and steady propagation fields.
We also study the steady propagation of negative streamers in different background fields.
In \sref{sec:discussion and analysis}, we address several important questions about negative streamers, and compare their properties with those of positive streamers.
Finally, we summarize our findings in \sref{sec:conclusions and outlook}, and provide some ideas for future studies.

\section{Simulation model}\label{sec:simulation model}

We use a 2D axisymmetric drift-diffusion-reaction type fluid model with the local field approximation to simulate single negative streamers in artificial dry air, consisting of 80\% $\mathrm{N}_2$ and 20\% $\mathrm{O}_2$, at 300 Kelvin and 1 bar.
We use \verb"Afivo-streamer" \cite{teunissen2017}, an open-source code for streamer fluid simulations.
The code is based on the Afivo framework~\cite{teunissen2018}, which includes adaptive mesh refinement (AMR), shared-memory (OpenMP) parallelism and a geometric multigrid solver.
This code was compared to five other fluid simulation codes in the comparison study of~\cite{bagheri2018}.
Furthermore, simulations using \verb"Afivo-streamer" were recently compared against experiments in~\cite{li2021a}, and compared against particle-in-cell simulations both in 2D and 3D for positive streamers in air in~\cite{wang2022}, which found good quantitative agreement.

\subsection{Model equations}\label{sec:model equations}

The temporal evolution of the electron density $n_{\mathrm e}$ is given by
\begin{equation}\label{eq:evolution of ne}
    \partial_t n_{\mathrm e} = \nabla \cdot (\mu_{\mathrm e} \boldsymbol{\mathrm E} n_{\mathrm e} + D_{\mathrm e} \nabla n_{\mathrm e}) + S_{\mathrm i} - S_{\mathrm{att}} + S_{\mathrm{det}} + S_{\mathrm{ph}}\,,
\end{equation}
where $\mu_{\mathrm e}$ is the electron mobility, $\boldsymbol{\mathrm E}$ the electric field, $D_{\mathrm e}$ the electron diffusion coefficient, and $S_{\mathrm i}$, $S_{\mathrm{att}}$, $S_{\mathrm{det}}$ and $S_{\mathrm{ph}}$ are the source terms for electron impact ionization, electron attachment, electron detachment and non-local photoionization, respectively.
The densities of other species involved in the model evolve according to the reactions listed in \tref{tab:list of reactions}.
Ions and neutrals are assumed to be immobile due to their much larger mass than that of electrons.
We therefore do not include ion motion in the model, as is discussed in \ref{sec:ion motion}.

\begin{table*}
\centering
\captionsetup{width=0.92\textwidth}
\caption{\label{tab:list of reactions}List of reactions included in the model, with reaction rate coefficients and references.
The symbol M denotes a neutral molecule (either $\mathrm N_2$ or $\mathrm O_2$).
$E/N$ is the reduced electric field.
% $k_{11}$ and $k_{12}$ are quenching rate constants for collisions with $\mathrm N_2$ and $\mathrm O_2$, respectively.
% The radiative lifetime of N$_2(C^3\Pi_u) is $1/ k_{13} =$ 42\,ns.
}
\begin{tabular*}{0.92\textwidth}{l@{\extracolsep{\fill}}llc}
\br
No. & Reaction & Reaction rate coefficient & Reference\\
\mr
\multicolumn{4}{l}{Electron impact ionization}\\
R1 & $\mathrm e + \mathrm N_2 \to \mathrm e + \mathrm e + \mathrm N_2^+$ (15.60\,eV) & $k_1(E/N)$ & \cite{Phelps1985}\\
R2 & $\mathrm e + \mathrm N_2 \to \mathrm e + \mathrm e + \mathrm N_2^+$ (18.80\,eV) & $k_2(E/N)$ & \cite{Phelps1985}\\
R3 & $\mathrm e + \mathrm O_2 \to \mathrm e + \mathrm e + \mathrm O_2^+$ (12.06\,eV) & $k_3(E/N)$ & \cite{Phelps1985}\\
\mr
\multicolumn{4}{l}{Electron attachment}\\
R4 & $\mathrm e + \mathrm O_2 + \mathrm O_2 \to \mathrm O_2^- + \mathrm O_2$ & $k_4(E/N)$ & \cite{Phelps1985}\\
R5 & $\mathrm e + \mathrm O_2 \to \mathrm O^- + \mathrm O$ & $k_5(E/N)$ & \cite{Phelps1985}\\
\mr
\multicolumn{4}{l}{Electron detachment}\\
R6 & $\mathrm O_2^- + \mathrm M \to \mathrm e + \mathrm O_2 + \mathrm M$ & $k_{6}=1.24\times10^{-17}\exp(-(\frac{179}{8.8+E/N})^2)\,\mathrm{m^3\,s^{-1}}$ & \cite{pancheshnyi2013}\\
R7 & $\mathrm O^- + \mathrm N_2 \to \mathrm e + \mathrm{N_2O}$ & $k_{7}=1.16\times10^{-18}\exp(-(\frac{48.9}{11+E/N})^2)\,\mathrm{m^3\,s^{-1}}$ & \cite{pancheshnyi2013}\\
\mr
\multicolumn{4}{l}{Negative ion conversion}\\
R8 & $\mathrm O^- + \mathrm O_2 \to \mathrm O_2^- + \mathrm O$ & $k_{8}=6.96\times10^{-17}\exp(-(\frac{198}{5.6+E/N})^2)\,\mathrm{m^3\,s^{-1}}$ & \cite{pancheshnyi2013}\\
R9 & $\mathrm O^- + \mathrm O_2 + \mathrm M \to \mathrm O_3^- + \mathrm M$  & $k_{9}=1.10\times10^{-42}\exp(-(\frac{E/N}{65})^2)\,\mathrm{m^6\,s^{-1}}$ & \cite{pancheshnyi2013}\\
\mr
\multicolumn{4}{l}{Electron excitation}\\
R10 & $\mathrm e + \mathrm N_2 \to \mathrm e + \mathrm{N}_2(C^3\Pi_u)$ & $k_{10}(E/N)$ & \cite{Phelps1985}\\
\mr
\multicolumn{4}{l}{Quenching}\\
R11 & $\mathrm{N}_2(C^3\Pi_u) + \mathrm N_2 \to \mathrm N_2 + \mathrm N_2$ & $k_{11}=1.3\times10^{-17}\,\mathrm{m^3\,s^{-1}}$ & \cite{pancheshnyi2005c}\\
R12 & $\mathrm{N}_2(C^3\Pi_u) + \mathrm O_2 \to \mathrm O_2 + \mathrm N_2$ & $k_{12}=3.0\times10^{-16}\,\mathrm{m^3\,s^{-1}}$ & \cite{pancheshnyi2005c}\\
\mr
\multicolumn{4}{l}{Radiation}\\
R13 & $\mathrm{N}_2(C^3\Pi_u) \to \mathrm{N}_2(B^3\Pi_g) + h \nu$ & $k_{13}=1/(42\,\mathrm{ns})$ & \cite{pancheshnyi2005c}\\
\br
\end{tabular*}
\end{table*}

The electric field $\boldsymbol{\mathrm E}$ is calculated in the electrostatic approximation as
\begin{equation}
    \boldsymbol{\mathrm E} = - \nabla \phi\,,
\end{equation}
where the electrostatic potential $\phi$ is obtained by solving Poisson's equation
\begin{equation}
\label{eq:Poisson equation}
    \nabla^2 \phi = - \frac{\rho}{\varepsilon_0}\,,
    % \quad
    % \rho = e\,(n_{\mathrm i}^+ - n_{\mathrm i}^- - n_{\mathrm e})\,,
\end{equation}
where $\rho$ is the space charge density and $\varepsilon_0$ is the vacuum permittivity.
% $e$ the elementary charge, and $n_{\mathrm i}^+$ and $n_{\mathrm i}^-$ are the total number density of all positive and negative ion species, respectively.

\subsection{Chemical reactions and input data}\label{sec:reactions and input data}

We use a relatively small set of chemical reactions, which are listed in \tref{tab:list of reactions}. % which contains:  electron impact ionization (R1--R3), electron attachment (R4, R5), electron detachment (R6, R7), negative ion conversion (R8, R9), electron excitation (R10), quenching (R11, R12) and radiation (R13). 
Note that positive ion conversion and electron-ion recombination are not included. 
Such reactions can play an important role in air, for example due to the formation of O$_4^+$ with which electrons can dissociatively recombine, but in \ref{sec:recombination time} we show that they have a relatively minor effect on the negative streamers studied here.
From \tref{tab:list of reactions}, the source terms for electron impact ionization, attachment and detachment are computed as
\begin{equation}
\label{eq:impact ionization source term}
    S_{\mathrm i} = k_1 n_{\mathrm e} [\mathrm N_2] + k_2 n_{\mathrm e} [\mathrm N_2] + k_3 n_{\mathrm e} [\mathrm O_2]\,,
\end{equation}
\begin{equation}
    S_{\mathrm{att}} = k_4 n_{\mathrm e} [\mathrm O_2]^2 + k_5 n_{\mathrm e} [\mathrm O_2]\,,
\end{equation}
\begin{equation}
    S_{\mathrm{det}} = k_6 [\mathrm M] [\mathrm O_2^-] + k_7 [\mathrm N_2] [\mathrm O^-]\,,
\end{equation}
where $k_i$ ($i = 1, 2, \dots, 7$) are the reaction rate coefficients, $[\mathrm N_2]$, $[\mathrm O_2]$, $[\mathrm O^-]$ and $[\mathrm O_2^-]$ are the species densities, and $[\mathrm M] = [\mathrm N_2] + [\mathrm O_2]$.
We assume that the number density of $[\mathrm M]$ does not change in the model.
Note that the three-body attachment reaction, $\mathrm e + \mathrm{O_2 + N_2} \to \mathrm{O_2^- + N_2}$, is not included in the model because its reaction rate coefficient is about two to three orders of magnitude smaller than $k_4$~\cite{kossyi1992}.

Non-local photoionization in air occurs ahead of a streamer discharge when an UV photon ionizes an oxygen molecule at some isotropically distributed distance.
The UV photon is emitted from an excited nitrogen molecule with a wavelength in the range 98--102.5\,nm.
Photoionization is typically considered as an important source of free electrons for both positive and negative streamers in air~\cite{babaeva1997, yi2002, liu2004, luque2007, luque2008, naidis2009, starikovskiy2020}.
Here we use Zheleznyak's model to describe photoionization~\cite{zheleznyak1982}.
Then the photoionization source term $S_{\mathrm{ph}}$ in equation \eref{eq:evolution of ne} is given by
\begin{equation}
\label{eq:photoionization source term}
    S_{\mathrm{ph}}(\boldsymbol{r}) = \int \frac{I(\boldsymbol{r}')f(\left| \boldsymbol{r} - \boldsymbol{r}' \right|)}{4\pi \left| \boldsymbol{r} - \boldsymbol{r}' \right |^2}\,\mathrm{d}^3 r'\,,
\end{equation}
where $\boldsymbol{r}$ is a given observation point, $\boldsymbol{r}'$ the source point emitting UV photons, $I(\boldsymbol{r}')$ the source of ionizing photons, $f(\left| \boldsymbol{r} - \boldsymbol{r}' \right|)$ the photon absorption function, and $4\pi \left| \boldsymbol{r}-\boldsymbol{r}' \right|^2$ is a geometric factor.
In Zheleznyak's model, $I(\boldsymbol{r}')$ is proportional to the electron impact ionization source term $S_{\mathrm i}$ given by equation \eref{eq:impact ionization source term} as
\begin{equation}
    I(\boldsymbol{r}') = \frac{p_q}{p+p_q} \xi S_{\mathrm{i}}\,,
\end{equation}
where $p_q=$ 40\,mbar is the quenching pressure, $p$ the gas pressure, and $\xi$ is a proportionality factor, which is here set to $\xi=$ 0.075 for simplicity as in~\cite{bagheri2019, francisco2021a, francisco2021e, li2021a, wang2022}, although it is in principle field-dependent~\cite{zheleznyak1982}.
% Only include this in PhD thesis
% Furthermore, $f(\left| \boldsymbol{r} - \boldsymbol{r}' \right|)$ evaluating the probability of photon absorption can be rewritten considering $\Delta r = \left| \boldsymbol{r} - \boldsymbol{r}' \right|$ as
% \begin{equation}
%     f(\Delta r) = \frac{\mathrm{exp}(-\chi_{\mathrm{min}} p_{\mathrm O_2} \Delta r) - \mathrm{exp}(-\chi_{\mathrm{max}} p_{\mathrm O_2} \Delta r)}{\Delta r \ln(\chi_{\mathrm{max}}/\chi_{\mathrm{min}})}\,,
% \end{equation}
% where $\chi_{\mathrm{max}}$\,$\approx$\,1.5\,$\times 10^2$\,/(mm\,bar), $\chi_{\mathrm{min}}$\,$\approx$\,2.6\,/(mm\,bar), and $p_{\mathrm{O_2}}$ is the partial pressure of oxygen molecules.
We solve equation \eref{eq:photoionization source term} using the so-called Helmholtz approximation~\cite{luque2007, bourdon2007}, for which the absorption function is computed from Bourdon's three-term expansion~\cite{bourdon2007}.
See the appendix of~\cite{bagheri2018} for more information.

With the local field approximation, the electron velocity distribution is assumed to be relaxed to the local electric field.
Therefore the transport coefficients $\mu_{\mathrm e}$ and $D_{\mathrm e}$, and the reaction rate coefficients ($k_1$\,--\,$k_{10}$) are functions of the reduced electric field $E/N$, where $E$ is the electric field and $N$ is the gas number density.
These coefficients were computed with BOLSIG$+$, a two-term electron Boltzmann equation solver~\cite{hagelaar2005}, using the temporal growth model.
Electron-neutral scattering cross sections for $\mathrm N_2$ and $\mathrm O_2$ were obtained from the Phelps database~\cite{Phelps1985, Phelps_database}.
% Only include this in PhD thesis
% Furthermore, these coefficients were tabulated at regularly spaced electric field strengths using 1000 steps and were interpolated linearly in the model.
% Note that even though using a very fine grid and a small time step, some numerical oscillations in the streamer properties may still appear, which are caused by the interpolation errors using few number of points in the tabulated input data~\cite{bagheri2018, bagheri2020}. Therefore, a high resolution input data or a higher order of the interpolating function, e.g., cubic spline interpolation, is required to remove the numerical oscillations.

\subsection{Computational domains and initial conditions}\label{sec:simulation domain}

\begin{figure}
    \centering
    \includegraphics[width=0.49\textwidth]{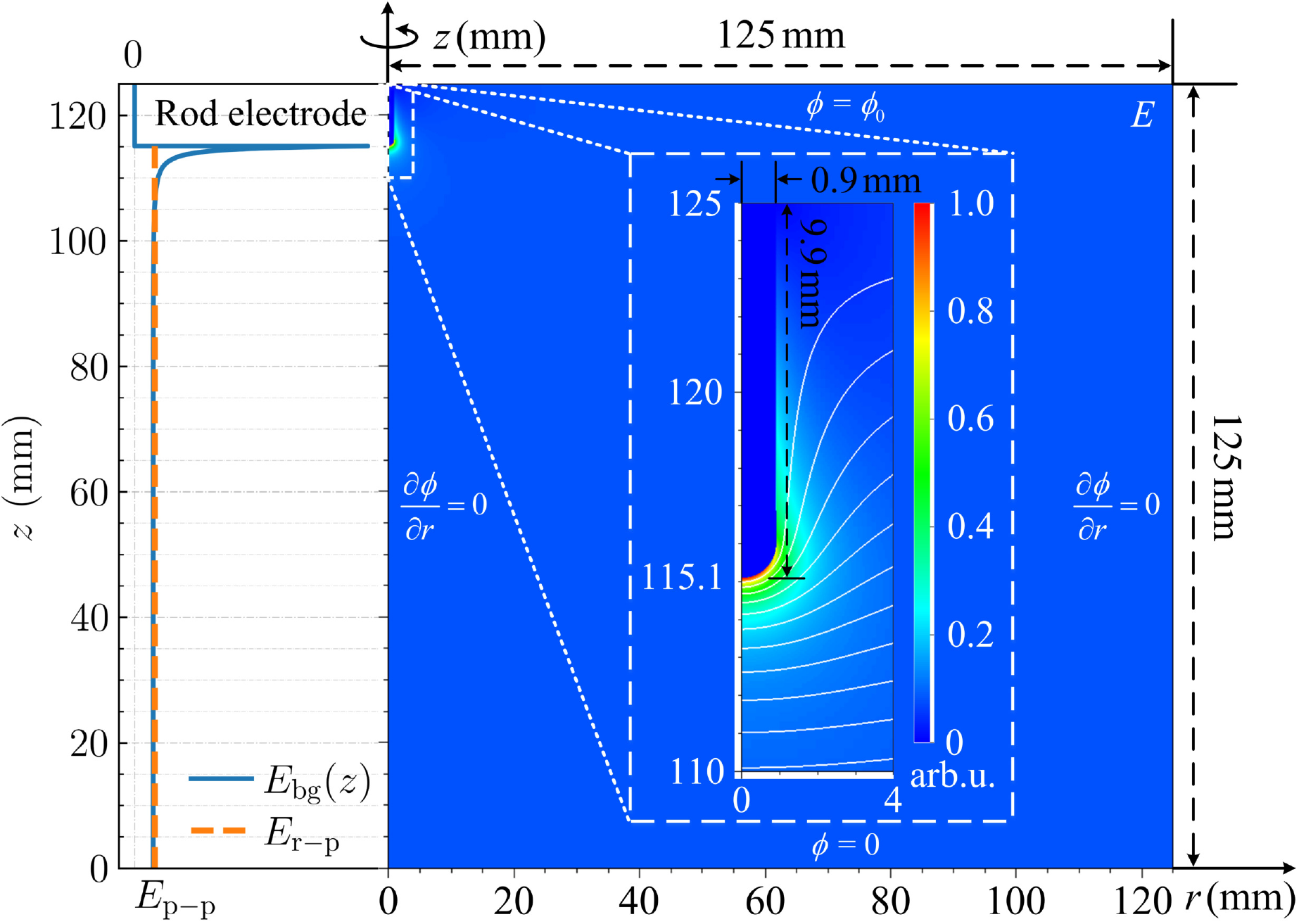}
    \caption{Schematic view of computational domain A.
    Right: the rod electrode geometry, boundary conditions for the electric potential $\phi$ and  initial electric field $E$ without a streamer.
    Left: the axial background electric field $E_\mathrm{bg}(z)$ and the average electric field $E_{\mathrm{r-p}}$ between the rod electrode tip and the grounded electrode.
    $E_{\mathrm{p-p}}$ is the average electric field between the upper and lower plate electrodes.}
    \label{fig:simulation domain}
\end{figure}

% We use two axisymmetric computational domains in the simulations, as described in \tref{tab:computational domain}.
% For these two computational domains, we both implement a rod-shaped electrode with a semi-spherical tip at the center of the upper plate electrode.
% Except for the domain size and the rod electrode geometry, initial and boundary conditions are the same.
% Domain A and its boundary conditions are further illustrated in \fref{fig:simulation domain}.
We use two axisymmetric computational domains in the simulations, as described in \tref{tab:computational domain}.
Domain A measures 125\,mm in the $r$ and $z$ directions, whereas domain B measures 300\,mm in both directions.
Both domains have a plate-plate geometry with a rod-shaped electrode of length $L_\mathrm{rod}$ placed at the center of the upper plate, see figure \ref{fig:simulation domain}.
For domain A we use $L_\mathrm{rod}=9.9$\,mm and for domain B we use ten different electrode lengths ranging from 3.3\,mm to 26.4\,mm.
All electrodes have a semi-spherical tip and a radius $R_\mathrm{rod}$ given by $L_\mathrm{rod}$/11, which means that longer electrodes are also wider.
Also note that for longer electrodes, the distance between the electrode tip and the grounded electrode is shorter, but the electrode length is always significantly smaller than the domain length, see figure \ref{fig:simulation domain}.

\begin{table*}
\centering
\captionsetup{width=0.85\textwidth}
\caption{\label{tab:computational domain} Description of the two axisymmetric computational domains used in the present paper. $z_{\mathrm{head}}$ is the streamer position at which the electric field has a maximum, see \sref{sec:definitions}.}
\begin{tabular*}{0.85\textwidth}{l @{\extracolsep{\fill}} l l}
%\begin{tabular}{c c c}
\br
 & Domain A & Domain B\\
\mr
Domain size $(r, z)$ & 125\,mm, 125\,mm & 300\,mm, 300\,mm\\
Rod electrode & One fixed geometry & Ten different geometries\\
Rod electrode length $L_\mathrm{rod}$ & 9.9\,mm & 3.3\,mm to 26.4\,mm\\
Rod electrode radius $R_\mathrm{rod}$ & 0.9\,mm & $L_\mathrm{rod}/11$\\
Simulations stop when & $z_{\mathrm{head}} = 30$\,mm or fading & $z_{\mathrm{head}} = 100$\,mm or fading\\
Simulation results & sections \ref{sec:three examples}, \ref{sec:stab-fields-steady} and \ref{sec:fading streamers} & \sref{sec:steady streamers}\\
\br
\end{tabular*}
\end{table*}

We apply the same initial and boundary conditions for both domains.
A fixed electric potential $\phi_0$ is applied on the upper domain boundary and the rod electrode.
The lower domain boundary is grounded. A homogeneous Neumann boundary condition for the electric potential is applied on the outer axial boundary.

For the electron density, homogeneous Neumann boundary conditions are applied on all domain boundaries, including the rod electrode.
Secondary electron emission due to ions and photons is not included.
It would be more realistic to only allow an outflow of electrons from the rod electrode due to secondary emission~\cite{babaeva2016a, zhang2021a}.
However, this would give rise to very high electric fields around the rod electrode and steep density gradients, which are numerically challenging to simulate.
The artificial boundary conditions used here affect discharge inception, and therefore also the initial streamer properties.
However, we here focus on steady discharge evolution far from electrodes.
For most of the cases we study the back part of the discharge has a low conductivity (see e.g.\ figure \ref{fig:ne for ten steady streamers}), so the artifical boundary conditions should not have a major effect on this steady propagation.

The axial background electric field $E_\mathrm{bg}(z)$ in \fref{fig:simulation domain} is almost uniform, except for a small region near the rod electrode.
Far away from the rod electrode, $E_\mathrm{bg}(z)$ is almost equal to the average electric field $E_{\mathrm{p-p}}$ between the upper and lower plate electrodes.
In addition, there is the average electric field $E_{\mathrm{r-p}}$ between the rod electrode tip and the grounded electrode, which is about 1.086\,$E_{\mathrm{p-p}}$ in domain A.
Note that this ratio varies for different high-voltage electrode geometries used in domain B, as will be shown in \fref{fig:ten steady streamers}.

As an initial condition, homogeneous background ionization with a density of $10^{11}\,\mathrm m^{-3}$ for both electrons and positive ions (N$_2^+$) is included.
All other ion densities are initially zero.

\subsection{Afivo AMR framework}

The open-source Afivo framework~\cite{teunissen2018} is used in the model to provide AMR for computational efficiency.
The refinement criteria are given by
\begin{itemize}
    \item Refine if $\Delta x > c_0 c_1 / \alpha(c_1 E)$\,,
    \item De-refine if $\Delta x < \mathrm{min}\{0.125\, c_0 c_1/\alpha(c_1 E),\,d_0\} $\,,
    \item $\Delta x \leqslant$\,1.0\,mm\,,
\end{itemize}
where $\Delta x$ is the grid spacing, which is identical in all directions, and $\alpha(E)$ is the field-dependent ionization coefficient calculated as $S_{\mathrm i}/(\mu_{\mathrm e} E n_{\mathrm e})$ from equation (\ref{eq:impact ionization source term}).
We use $c_0=0.8$ and $c_1=1.25$ to balance the refinement ahead and on the sides of the streamer, and $d_0=0.1$\,mm.
The grid has a minimal size of $\Delta x_\mathrm{min}=$ 1.8\,$\mu$m in the simulations.

The Afivo framework provides a geometric multigrid solver to efficiently solve Poisson's equation \eref{eq:Poisson equation} and the Helmholtz equation \eref{eq:photoionization source term}.
The rod electrode is implemented by modifying the multigrid method with a level-set function~\cite{teunissen2022}.
% We apply a fixed electric potential on the rod electrode as a boundary condition, which is implemented by modifying the multigrid methods using a level-set function.

\subsection{Definitions}\label{sec:definitions}

We refer to the moment we apply the high voltage $V$ as $t=$ 0\,ns.
\textbf{The streamer position} $z_{\mathrm{head}}$ is defined as the axial location at which the electric field has a maximum.
\textbf{The streamer length} $L_\mathrm{s}$ is then computed as $L_\mathrm{s} = z_0 - z_{\mathrm{head}}$\,, where $z_0$ is the axial location of the rod electrode tip.
\textbf{The streamer velocity} $v$ is computed as the numerical time derivative of the streamer position, measured every $0.5$\,ns.
This causes some fluctuations, so we use a second order Savitzky--Golay filter with a window width of 11 to smooth the velocity.

\textbf{The head potential} $\delta \phi$ is here defined as the potential difference induced by the streamer at its head: $\delta \phi$ = $\phi(z_\mathrm{head}, t) - \phi(z_\mathrm{head}, 0)$, where $\phi(z_\mathrm{head}, t)$ and $\phi(z_\mathrm{head}, 0)$ are the actual electric potential and background electric potential at the streamer head, respectively.

\textbf{The maximal electron drift velocity} $v_\mathrm{dmax}$ is here defined as the electron drift velocity corresponding to the maximal electric field at a particular instant of time.

\textbf{The background electric field} $E_{\mathrm{p-p}}$ is here defined as the average electric field between the upper and lower plate electrodes.
\textbf{The average electric field} $E_{\mathrm{r-p}}$ is measured between the rod electrode tip and the grounded electrode.

\textbf{The breakdown field} $E_\mathrm{k}$ is defined at which the impact ionization rate is equal to the attachment rate.
With our transport data, $E_\mathrm{k}=28$\,kV/cm in air at 300\,K and 1 bar.

For a negative streamer fading out in a weak background electric field, \textbf{streamer fading} is here arbitrarily defined to occur when the maximal electric field $E_{\mathrm{max}}$ decreases to 1.25$E_\mathrm{k}$ (35\,kV/cm).
Then the streamer position and length at this moment are defined as \textbf{the fading position} $z_{\mathrm{s}}$ and \textbf{the maximal streamer length} $L_\mathrm{smax}$, respectively.
We will further discuss the definition of negative streamer fading in \sref{sec:definition of fading}.

To quantitatively compare with experiments, the streamer diameter $d$ is here defined as the full width at half maximum (FWHM) \textbf{optical diameter} $d_{\mathrm{optical}}$.
We compute the optical diameter from the $\mathrm{N}_2(C^3\Pi_u)$ density, which is approximately proportional to the light emission intensity, as most light comes from the $\mathrm{N}_2(C^3\Pi_u) \to \mathrm{N}_2(B^3\Pi_g)$ transition~\cite{pancheshnyi2000}.
First a forward Abel transform is applied~\cite{Hansen1985}, after which the emission is vertically integrated to obtain a radial profile, and then the FWHM is determined.
A more detailed description is given in~\cite{li2021a}.
% in cylindrical coordinates to obtain 2D projection of the $\mathrm{N}_2(C^3\Pi_u)$ density in Cartesian coordinates.
% From the 2D projection, the area near the rod electrode with high $\mathrm{N}_2(C^3\Pi_u)$ density was removed to more accurately reflect light emission at the streamer head.
% Then the $\mathrm{N}_2(C^3\Pi_u)$ density was integrated vertically to obtain its 1D profile along the horizontal axis, of which the FWHM value is the optical diameter.

There are actually several different definitions of the streamer diameter, which lead to a different value~\cite{babaeva1996, pancheshnyi2005c}.
To compare different definitions of the streamer diameter, two electrodynamic definitions of the streamer diameter are introduced in \ref{sec:diameter comparison}, namely $d_{Ez}$, related to the decay of the electric field ahead of the streamer, and $d_{Er}$, related to the location of the maximal radial electric field.

\section{Simulation results}\label{sec:simulation results}

In \sref{sec:three examples}, we first present examples of accelerating, steady and fading negative streamers in air.
Then we discuss stability fields and steady propagation fields in \sref{sec:stab-fields-steady}.
Next, the dependence of negative streamer fading on the applied voltage is studied in \sref{sec:fading streamers}.
Finally, we investigate steady negative streamers in different background fields in \sref{sec:steady streamers}.

\subsection{Three distinct evolutions of negative streamers}\label{sec:three examples}

\begin{figure*}
    \centering
    \includegraphics[width=0.95\textwidth]{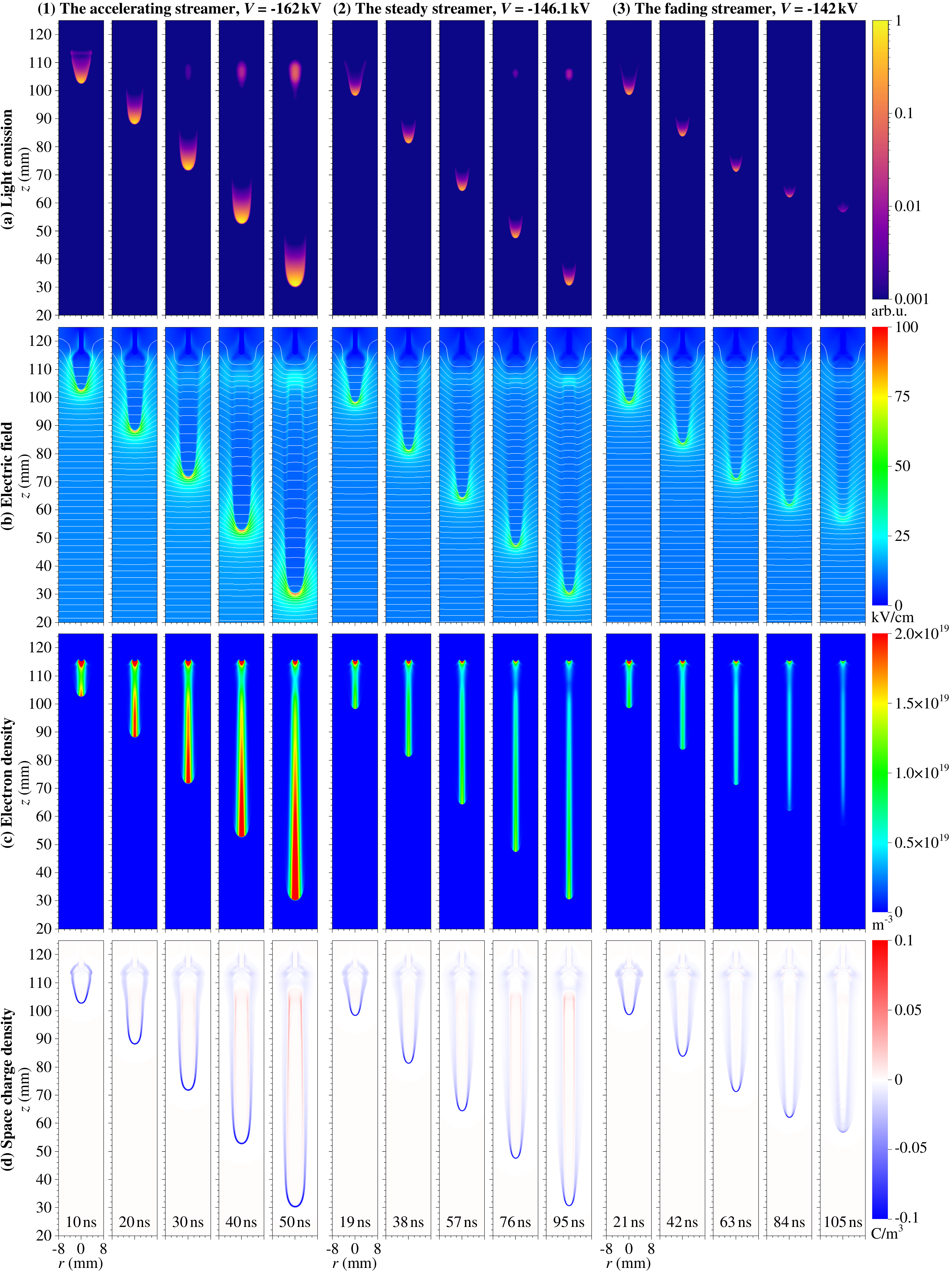}
    \caption{Time evolution of (a) the light emission profile, (b) the electric field $E$ with white equipotential lines, (c) the electron density $n_{\mathrm e}$ and (d) the space charge density $\rho$ for (1) accelerating (-162\,kV), (2) steady (-146.1\,kV) and (3) fading (-142\,kV) negative streamers in  air.
    The simulations were performed in domain A until $z_\mathrm{head}=30$\,mm or fading, see \tref{tab:computational domain}.
    Light emission was computed with a forward Abel transform, and results are shown using arbitrary units on a logarithmic scale.
    All panels are zoomed in into the region where -8 $\leqslant r \leqslant$ 8\,mm and 20 $\leqslant z \leqslant$ 125\,mm.}
    \label{fig:three examples evolution}
\end{figure*}

\begin{figure*}
    \centering
    \includegraphics[width=1\textwidth]{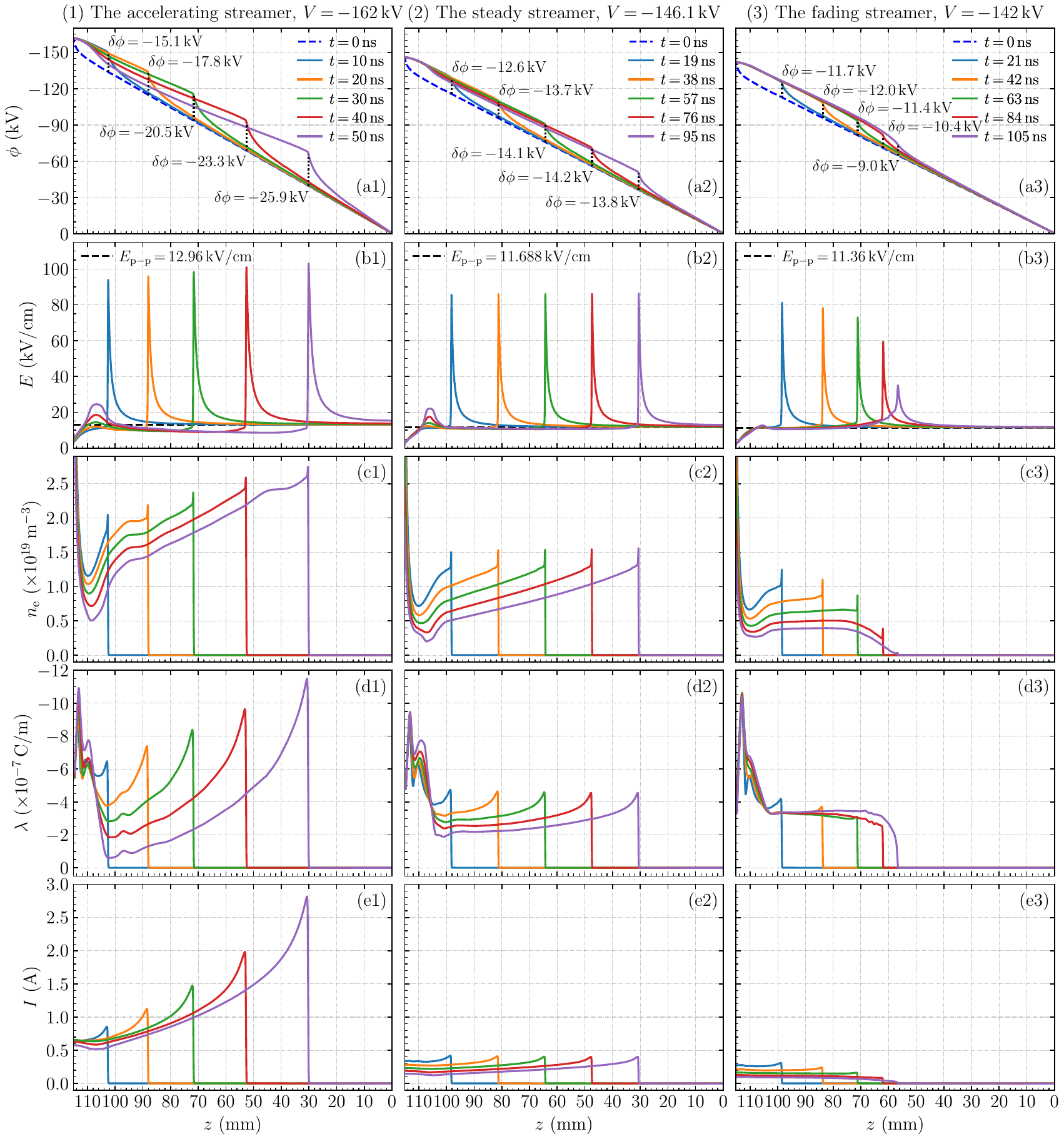}
    \caption{Time evolution of axial profiles, line charge density and electron conduction current for (1) accelerating ($-$162\,kV), (2) steady ($-$146.1\,kV) and (3) fading ($-$142\,kV) negative streamers corresponding to \fref{fig:three examples evolution}.
    The rows show: (a) the on-axis electric potential $\phi$ and the head potential $\delta \phi$ presented by vertical dotted lines, (b) the on-axis electric field $E$ and the background electric field $E_{\mathrm{p-p}}$ presented by dashed lines, (c) the on-axis electron density $n_{\mathrm e}$, (d) the line charge density $\lambda$, (e) the electron conduction current $I$.
    The quantities $\lambda$ and $I$ are computed by radially integrating the space charge density $\rho$ and the electron conduction current density ($j = e n_\mathrm{e} \mu_\mathrm{e} E$) up to $r=$ 15\,mm.
    All streamers propagate towards the right.}
    \label{fig:three examples profiles}
\end{figure*}

\begin{figure*}
  \centering
  \includegraphics[width=0.98\textwidth]{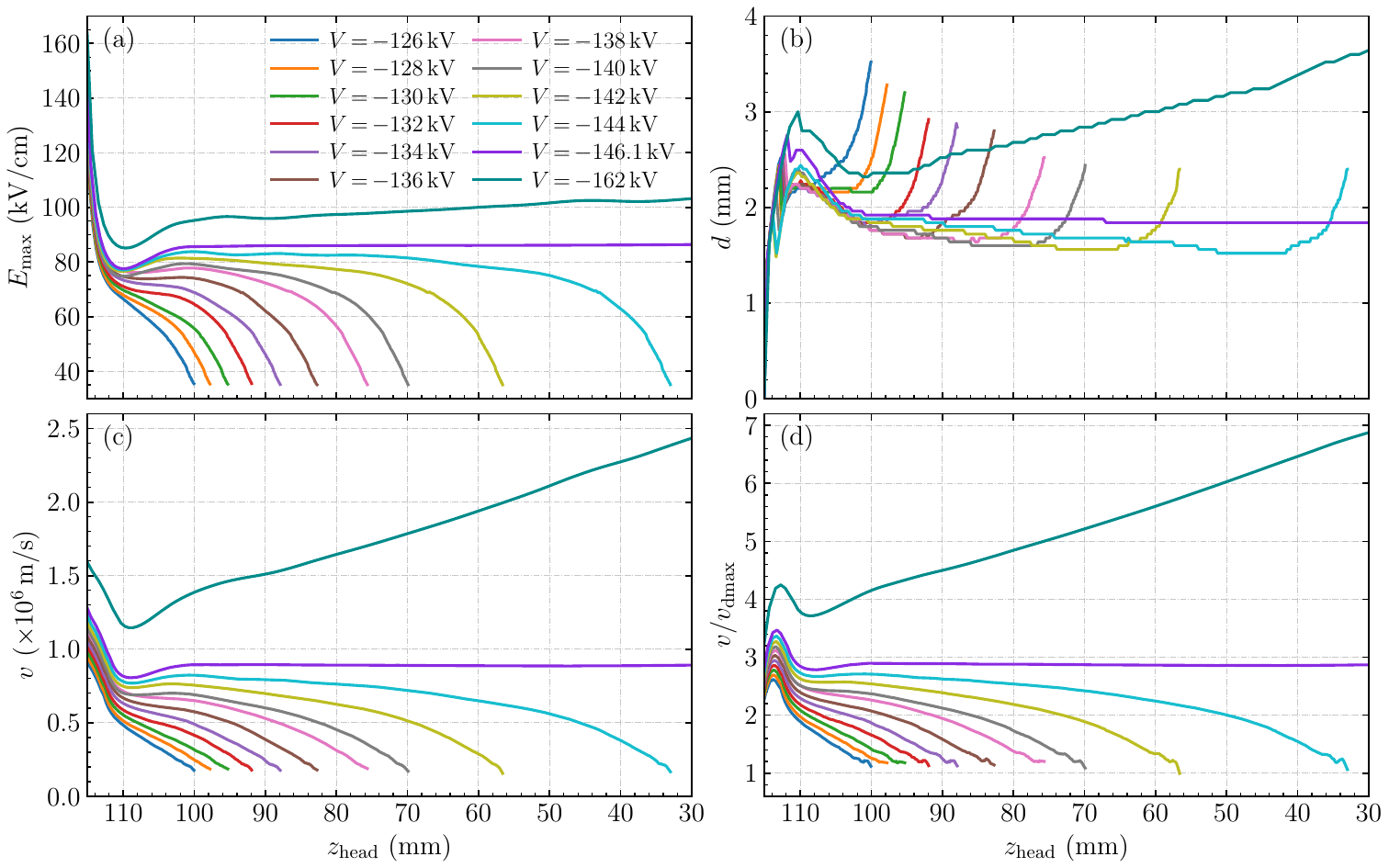}
  \caption{(a) The maximal electric field $E_{\mathrm{max}}$, (b) the optical diameter $d$, (c) the streamer velocity $v$ and (d) the ratio between streamer velocity and maximal electron drift velocity $v/v_{\mathrm{dmax}}$ versus the streamer position $z_{\mathrm{head}}$ for negative streamers in air.
  The simulations were performed in domain A until $z_\mathrm{head}=30$\,mm or fading, see \tref{tab:computational domain}, with applied voltages from $-$126\,kV to $-$162\,kV.
  All streamers propagate towards the right.}
  \label{fig:ten fading streamers}
\end{figure*}

In this section, we investigate negative streamer propagation in domain A, as shown in \tref{tab:computational domain}.
\Fref{fig:three examples evolution} shows examples of accelerating, steady and fading negative streamers in air at applied voltages of $-$162\,kV, $-$146.1\,kV and $-$142\,kV, which correspond to background fields ($E_{\mathrm{p-p}}$, see \sref{sec:definitions}) of 12.96\,kV/cm, 11.688\,kV/cm and 11.36\,kV/cm, respectively.
Axial profiles corresponding to these cases are shown in \fref{fig:three examples profiles}, and their evolutions of the maximal electric field, optical diameter and velocity versus the streamer position are shown in \fref{fig:ten fading streamers}.
Several differences can be observed between accelerating, steady and fading negative streamers, which are briefly discussed below.

\textbf{The accelerating streamer.}
With an applied voltage of $-$162\,kV the streamer accelerates.
\Fref{fig:ten fading streamers} shows that the streamer velocity and optical diameter increase almost linearly with streamer length, which are from about $1.4 \times 10^6$\,m/s to $2.4 \times 10^6$\,m/s and from about 2.4\,mm to 3.6\,mm, respectively.
In contrast, the maximal electric field is almost constant, with only a slight increase from about 95\,kV/cm to just above 100\,kV/cm.
\Fref{fig:three examples profiles} shows that the electron conduction current at the streamer head rapidly increases as the streamer grows, from about 0.9\,A to more than 2.5\,A.
Smaller increases are visible in the electron density and line charge density at the streamer head, which are from about $2.0 \times 10^{19} \, \mathrm{m}^{-3}$ to $2.7 \times 10^{19} \, \mathrm{m}^{-3}$ and from about $6.5 \times 10^{-7}$\,C/m to $11.5 \times 10^{-7}$\,C/m, respectively.
The streamer head is negatively charged and the streamer tail is positively charged.

The instantaneous light emission profiles resemble the streamer head shape, and the emission intensity increases with growing streamer length.
Note that there is also some light emission near the rod electrode.

The electric field inside the streamer channel is lower than the background electric field.
As the streamer grows there is a decrease in the internal electric field, to values as low as 8.6\,kV/cm when the streamer head is at $z_{\mathrm{head}}=$ 30\,mm.
This means that the head potential increases from about -15\,kV to -26\,kV, with the definition shown in \sref{sec:definitions}.

\textbf{The steady streamer.}
With a lower applied voltage of $-$146.1\,kV, the streamer propagates with a lower velocity of about $8.9 \times 10^5$\,m/s and all streamer properties stay nearly constant during its propagation.
\Fref{fig:three examples evolution} shows that the radius of the streamer head is smaller than the channel radius, both for the optical radius and electrodynamic radius, as defined in \sref{sec:definitions}.
This steady propagation mode is unstable, as will be shown in \sref{sec:steady streamers}.
It marks the transition between accelerating and decelerating streamers.
Quantities such as the maximal electric field, optical diameter, electron density, line charge density, electron conduction current and light emission intensity are all lower than those of the accelerating streamer.
However, the internal electric field is higher, namely about 11\,kV/cm.
This value can be compared to two definitions of a background field for the steady streamer, which are the background electric field $E_{\mathrm{p-p}}$ and the average electric field $E_{\mathrm{r-p}}$ between electrodes, see sections \ref{sec:simulation domain} and \ref{sec:definitions}.
First, the background electric field $E_{\mathrm{p-p}}$ is equal to
\begin{equation}
  \label{eq:background field}
  E_{\mathrm{p-p}} = |V|/d_\mathrm{plates} = 11.688\,\mathrm{kV/cm}\,,
\end{equation}
where $V = -146.1$\,kV is the applied voltage and $d_\mathrm{plates} = 125$\,mm is the distance between two plate electrodes.
Second, the average electric field $E_{\mathrm{r-p}}$ is given by
\begin{equation}
  \label{eq:average field}
  E_{\mathrm{r-p}} = |V|/(d_\mathrm{plates} - L_\mathrm{rod}) \approx 12.69\,\mathrm{kV/cm}\,,
\end{equation}
where $L_\mathrm{rod} = 9.9$\,mm is the length of the rod electrode, as shown in figure \ref{fig:simulation domain}.

\textbf{The fading streamer.}
When the voltage is further reduced to $-$142\,kV, the streamer first decelerates and then fades out.
The maximal electric field, electron density and electron conduction current at the streamer head decrease sharply with growing streamer length.
The internal electric field behind the streamer head is close to the background electric field of 11.36\,kV/cm. % but as the streamer fades out this internal field exceeds the background field over some length.
As the streamer grows the head potential decreases from about -12\,kV to -9\,kV, due to the voltage loss in the streamer channel.
The dominant propagation mechanism therefore gradually shifts from ionization to electron drift.
This radial outward drift increases the streamer diameter, see \fref{fig:ten fading streamers}, and it reduces the streamer's maximal electric field and electron density, leading to further deceleration.
When the maximal electric field approaches the breakdown field (see \sref{sec:definitions}), the discharge no longer generates significant ionization and light emission, as shown in \fref{fig:three examples evolution}.
The velocity is then comparable to the maximal electron drift velocity (see \sref{sec:definitions}), as will be further discussed in section \ref{sec:definition of fading}.
The remaining diffuse negative space charge prevents the inception of a new discharge from the high-voltage electrode.

The line charge density in the streamer head slightly decreases as the streamer grows.
One way to interpret this is that the electron conduction current in the channel cannot sustain its steady propagation.
However, when the streamer has faded, this current leads to an accumulation of negative charge near the streamer head, where the line charge density actually increases.
We will further investigate the fading of negative streamers in \sref{sec:fading streamers}.

\subsection{Stability fields and steady propagation fields}\label{sec:stab-fields-steady}

To understand whether a streamer is able to propagate in a certain background field, the phenomenological concept of a `stability field' has commonly been used.
There are actually several related definitions of such a stability field.
In experiments, the \textit{stability field} $E_\mathrm{st}$ is usually measured as $|V|/d$, where $V$ is the minimal applied voltage for which streamers can cross a gap of width $d$~\cite{phelps1971a, griffiths1976a, phelps1976, gallimberti1979a, allen1991, allen1995, allen1999, veldhuizen2002, briels2008, seeger2018, starikovskiy2020}.
Values of $E_\mathrm{st}$ between 10--12.5\,kV/cm for negative streamers in air were found in~\cite{Raizer1991, niemeyer1995, celestin2011, kochkin2014, starikovskiy2020}.
The values of $E_\mathrm{p-p}$ (11.688\,kV/cm) and $E_\mathrm{r-p}$ (12.69\,kV/cm) found for the steady propagation case in \sref{sec:three examples} agree well with these observations.

This concept of a stability field has also been used to estimate the maximal length $L_{\mathrm{smax}}$ (see \sref{sec:definitions}) of the streamers that do not cross the gap, using the following empirical equation~\cite{gallimberti1986, bujotzek2015, seeger2017}
\begin{equation}
  \label{eq:maximal length}
  \int_0^{L_{\mathrm{smax}}} (E_\mathrm{bg}(z)-E_\mathrm{st})\,\mathrm{d} z = 0\,,
\end{equation}
where $E_\mathrm{bg}(z)$ denotes the axial background electric field and the path from $z=0$ to $z=L_{\mathrm{smax}}$ corresponds to the streamer channel.
This means that a streamer stops when the average background electric field $\overline{E}_{\mathrm{bg}}$ over the streamer's length is equal to some fixed stability field $E_{\mathrm{st}}$.

The term stability field has also been used for the \textit{steady propagation field} corresponding to steady streamer propagation, as defined in \sref{sec:introduction}.
There have been several numerical studies of such steady propagation.
In~\cite{babaeva1997}, a negative streamer in air with a constant velocity was found in a background field of about 12.5\,kV/cm.
However, in~\cite{qin2014}, it was argued that such steady propagation fields can lie in a wide range, and steady negative streamers in air were found in background fields from 10\,kV/cm to 28\,kV/cm.
In \sref{sec:three examples}, we found a steady propagation field of 11.688\,kV/cm for negative streamers.
We will explore the variability of the steady propagation field in section \ref{sec:steady streamers}.

Steady propagation has also been observed for positive streamers in air.
In~\cite{francisco2021e}, a steady propagation field of 4.675\,kV/cm was found for positive streamers.
Later, Li \textit{et al} found steady propagation fields for positive streamers ranging from 4.1\,kV/cm to 5.5\,kV/cm~\cite{li2022}.
In~\cite{qin2014}, a larger range of steady propagation fields for positive streamers was found, from 4.4\,kV/cm to 20\,kV/cm.

\subsection{Dependence of negative streamer fading on the applied voltage}\label{sec:fading streamers}

We now investigate the dependence of negative streamer fading on the applied voltage in domain A as well, as shown in \tref{tab:computational domain}.
We consider ten applied voltages, evenly spaced between $-$126\,kV and $-$144\,kV.
\Fref{fig:ten fading streamers} shows the evolution of the maximal electric field, optical diameter, velocity and the ratio between streamer velocity and maximal electron drift velocity for the corresponding streamers until they have just faded, together with the accelerating and steady cases corresponding to \fref{fig:three examples evolution}.

With a lower applied voltage, fading occurs earlier.
However, the fading negative streamers are otherwise highly similar in terms of the temporal decay of the maximal electric field, velocity, optical diameter and head potential, as well as the spatial decay of the electron density, line charge density and electron conduction current along the streamer direction.
Before fading, the optical diameter of fading streamers remains approximately the same as that of the steady streamer.
During the fading phase, the streamer head becomes rather diffuse with the optical diameter increasing due to electron drift, and the streamer channel has lost most of conductivity due to attachment.
The variation in streamer velocity $v$ is larger than the variation in maximal electron drift velocity $v_\mathrm{dmax}$, see \sref{sec:definitions}, so that the variation in their ratio is mostly determined by the streamer velocity.
For the steady streamer at $V = -146.1$\,kV, the ratio $v/v_{\mathrm{dmax}}$ is about 2.87.
Note that for the fading streamers, this ratio decreases to about one at the moment of fading.
Since they keep propagating, the position of streamer fading and the corresponding maximal streamer length (see \sref{sec:definitions}) depend on some threshold for what is still considered a streamer.
We use an arbitrary fading criterion $E_{\mathrm{max}} \leqslant 35$\,kV/cm, see \sref{sec:definitions}.
If we would use another threshold, for example $E_{\mathrm{max}} \leqslant 50$\,kV/cm, this would lead to earlier fading.

\begin{figure}
    \centering
    \includegraphics[width=0.48\textwidth]{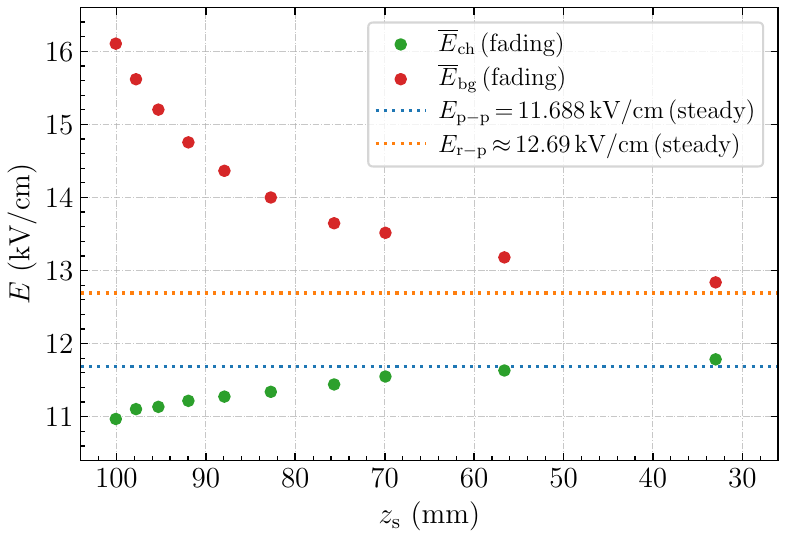}
    \caption{The average electric field $\overline{E}_{\mathrm{ch}}$ and the average background electric field $\overline{E}_{\mathrm{bg}}$, both measured along the streamer channel at the moment of fading for the ten fading streamers corresponding to \fref{fig:ten fading streamers}.
    The background electric field $E_{\mathrm{p-p}}$ and the average electric field $E_{\mathrm{r-p}}$ (between rod and plate electrode) for the steady streamer at $V=-146.1$\,kV are also shown.}
    \label{fig:stability field}
\end{figure}

We now look at the average electric field $\overline{E}_{\mathrm{ch}}$ and the average background electric field $\overline{E}_{\mathrm{bg}}$, both measured along the streamer channel at the moment of fading.
These fields are related to the stability field of fading streamers~\cite{babaeva1996, morrow1997, luque2014, nijdam2020a}.
\Fref{fig:stability field} shows $\overline{E}_{\mathrm{ch}}$ and $\overline{E}_{\mathrm{bg}}$ for the ten fading cases, as well as the steady case at $V=-146.1$\,kV.
In general, $\overline{E}_{\mathrm{bg}}$ is higher than $\overline{E}_{\mathrm{ch}}$.
Both $\overline{E}_{\mathrm{ch}}$ and $\overline{E}_{\mathrm{bg}}$ vary, in particular $\overline{E}_{\mathrm{bg}}$ for short streamers.
For long streamers, $\overline{E}_{\mathrm{bg}}$ decreases to about 12.8\,kV/cm, which is close to the average electric field $E_\mathrm{r-p}$ of 12.69\,kV/cm corresponding to steady propagation, as given by equation \eref{eq:average field}.
In contrast, $\overline{E}_{\mathrm{ch}}$ slightly increases with streamer length and it becomes close to the steady propagation field $E_{\mathrm{p-p}}$ of 11.688\,kV/cm corresponding to steady propagation, as given by equation \eref{eq:background field}.
We remark that $\overline{E}_{\mathrm{ch}}$ can be related to $\overline{E}_{\mathrm{bg}}$ via the head potential $\delta \phi$: $\delta \phi = (\overline{E}_{\mathrm{ch}} - \overline{E}_{\mathrm{bg}}) \times L_{\mathrm{smax}}$.

\begin{figure}
    \centering
    \includegraphics[width=0.48\textwidth]{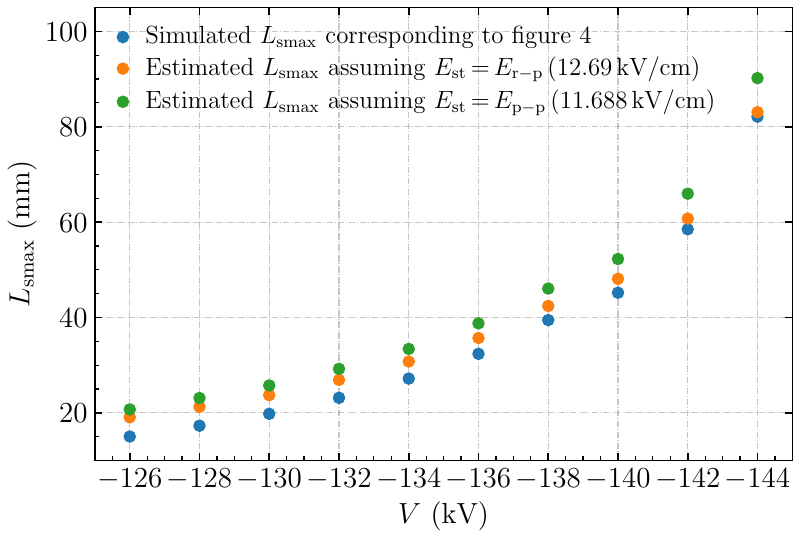}
    \caption{The simulated maximal streamer length $L_\mathrm{smax}$ versus the applied voltage $V$ for the ten fading streamers corresponding to \fref{fig:ten fading streamers}.
    Two estimates based on equation \eref{eq:maximal length} are also presented, by assuming $E_{\mathrm{st}}$ is equal to the value of $E_{\mathrm{p-p}}$ or $E_{\mathrm{r-p}}$ for the steady streamer at $V=-146.1$\,kV.}
    \label{fig:streamer length}
\end{figure}

We can estimate the maximal length $L_\mathrm{smax}$ of fading negative streamers using equation \ref{eq:maximal length}, by assuming the stability field $E_{\mathrm{st}}$ is equal to the value of $E_{\mathrm{p-p}}$ or $E_{\mathrm{r-p}}$ for the steady streamer at $V=-146.1$\,kV.
As shown in figure \ref{fig:streamer length}, these estimated maximal lengths are close to -- but slightly exceed -- the actual maximal lengths.
Note that differences depend not only on the value of $E_{\mathrm{st}}$, but also on the threshold used to identify negative streamer fading, and that the relative difference is smaller for longer streamers.
Our results confirm that the empirical equation \ref{eq:maximal length} can be used to estimate the maximal streamer length with some fixed conventional stability field.

\subsection{Dependence of steady propagation fields on the electrode geometry}\label{sec:steady streamers}

% In this section, we investigate how the high-voltage electrode geometry affects the properties of steady negative streamers in domain B, as shown in \tref{tab:computational domain}.
In this section, we investigate different steady negative streamers by changing the rod electrode geometry in domain B which has a length of 300\,mm, see \sref{sec:simulation domain} and \tref{tab:computational domain}.
We consider ten rod electrode lengths $L_\mathrm{rod}$, ranging from 3.3\,mm to 26.4\,mm, with the rod radius given by $L_\mathrm{rod}/11$.

\begin{figure*}
    \centering
    \includegraphics[width=1\textwidth]{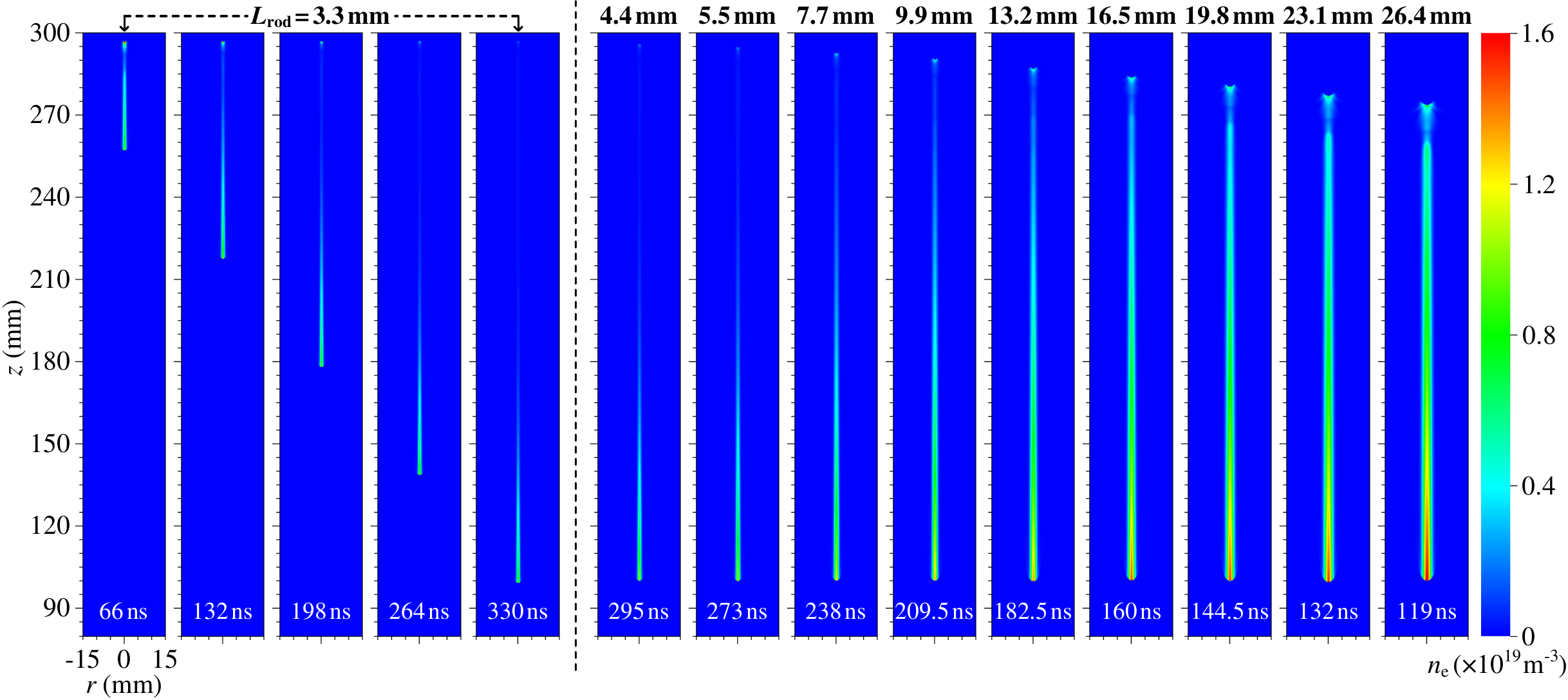}
    \caption{The electron density $n_{\mathrm e}$ for ten steady negative streamers in air with different rod electrode lengths $L_{\mathrm{rod}}$ from 3.3\,mm to 26.4\,mm.
    The simulations were performed in domain B until the streamer reached $z_\mathrm{head}=100$\,mm, see \tref{tab:computational domain}.
    The time evolution of $n_{\mathrm e}$ for the case of $L_{\mathrm{rod}}=3.3$\,mm is shown on the left.
    On the right, $n_{\mathrm e}$ is shown for the remaining nine steady streamers at $z_{\mathrm{head}}=100$\,mm.
    }
    \label{fig:ne for ten steady streamers}
\end{figure*}

\begin{figure*}
    \centering
    \includegraphics[width=0.98\textwidth]{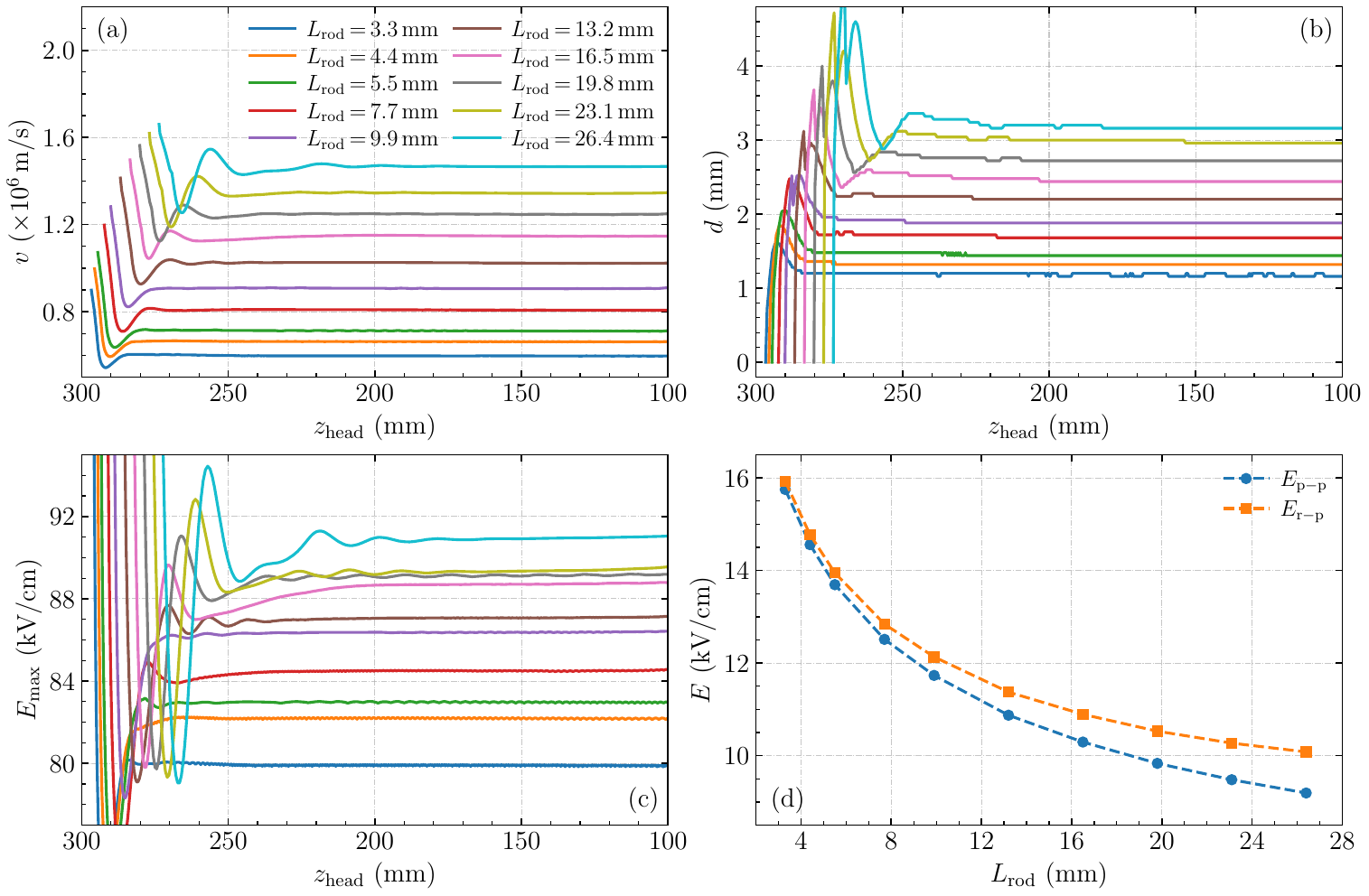}
    \caption{(a) The streamer velocity $v$, (b) the optical diameter $d$, and (c) the maximal electric field $E_{\mathrm{max}}$ versus the streamer position $z_{\mathrm{head}}$ for the ten steady streamers corresponding to \fref{fig:ne for ten steady streamers}.
    The corresponding background electric field $E_{\mathrm{p-p}}$ and the average electric field $E_{\mathrm{r-p}}$ between rod and plate electrode are shown in panel (d) versus the rod electrode length $L_{\mathrm{rod}}$.
    All streamers propagate towards the right.}
    \label{fig:ten steady streamers}
\end{figure*}

For each electrode geometry, the applied voltage was varied to find a steady negative streamer.
Only one steady solution was found for each electrode.
The electron density profiles of these ten steady streamers are shown in \fref{fig:ne for ten steady streamers}.
In \fref{fig:ten steady streamers}, the streamer velocity, optical diameter and maximal electric field are shown versus the streamer position, as well as the background electric field $E_{\mathrm{p-p}}$ and the average electric field $E_\mathrm{r-p}$.
Note that all streamer properties remain essentially constant.
With a longer rod electrode, the steady streamer propagates faster, and the optical diameter and maximal electric field are also higher, but the required background electric field is lower.

\begin{figure}
  \centering
  \includegraphics[width=0.48\textwidth]{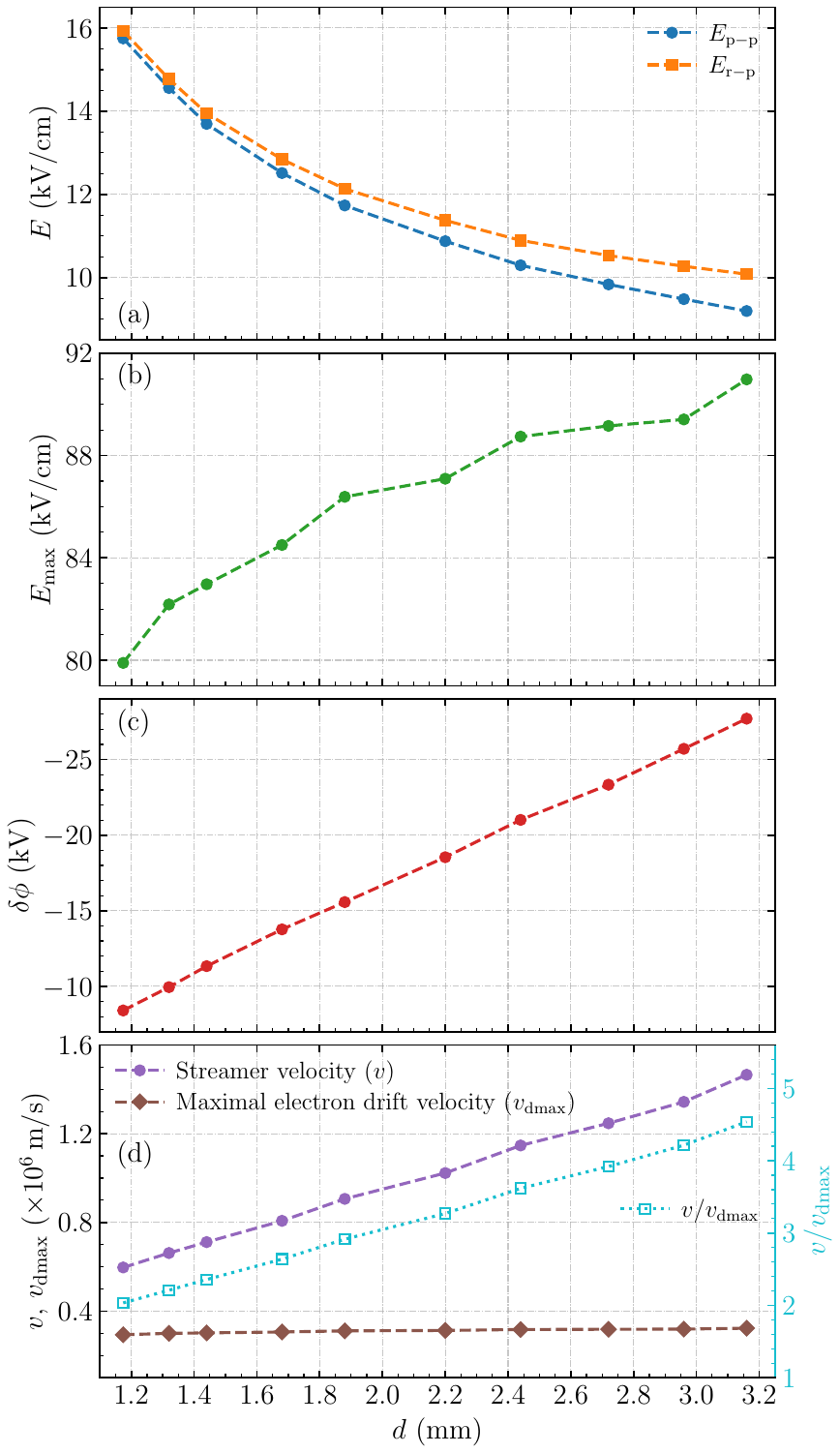}
  \caption{(a) The background electric field $E_{\mathrm{p-p}}$ and the average electric field $E_{\mathrm{r-p}}$, (b) the maximal electric field $E_{\mathrm{max}}$, (c) the head potential $\delta \phi$ and (d) the streamer velocity $v$, maximal electron drift velocity $v_{\mathrm{dmax}}$ and their ratio $v / v_{\mathrm{dmax}}$ versus the optical diameter $d$ for the ten steady streamers corresponding to \fref{fig:ne for ten steady streamers}.
  Each symbol in panels (b)--(d) represents one steady streamer, by taking an average over the propagation from $z_{\mathrm{head}}=$ 150\,mm to $z_{\mathrm{head}}=$ 100\,mm.}
  \label{fig:quantities versus d}
\end{figure}

The relationships between streamer properties and the optical diameter for these steady streamers are further illustrated in \fref{fig:quantities versus d}.
The streamer velocity and head potential have approximately linear relationships with the optical diameter.

\begin{figure}
    \centering
    \includegraphics[width=0.48\textwidth]{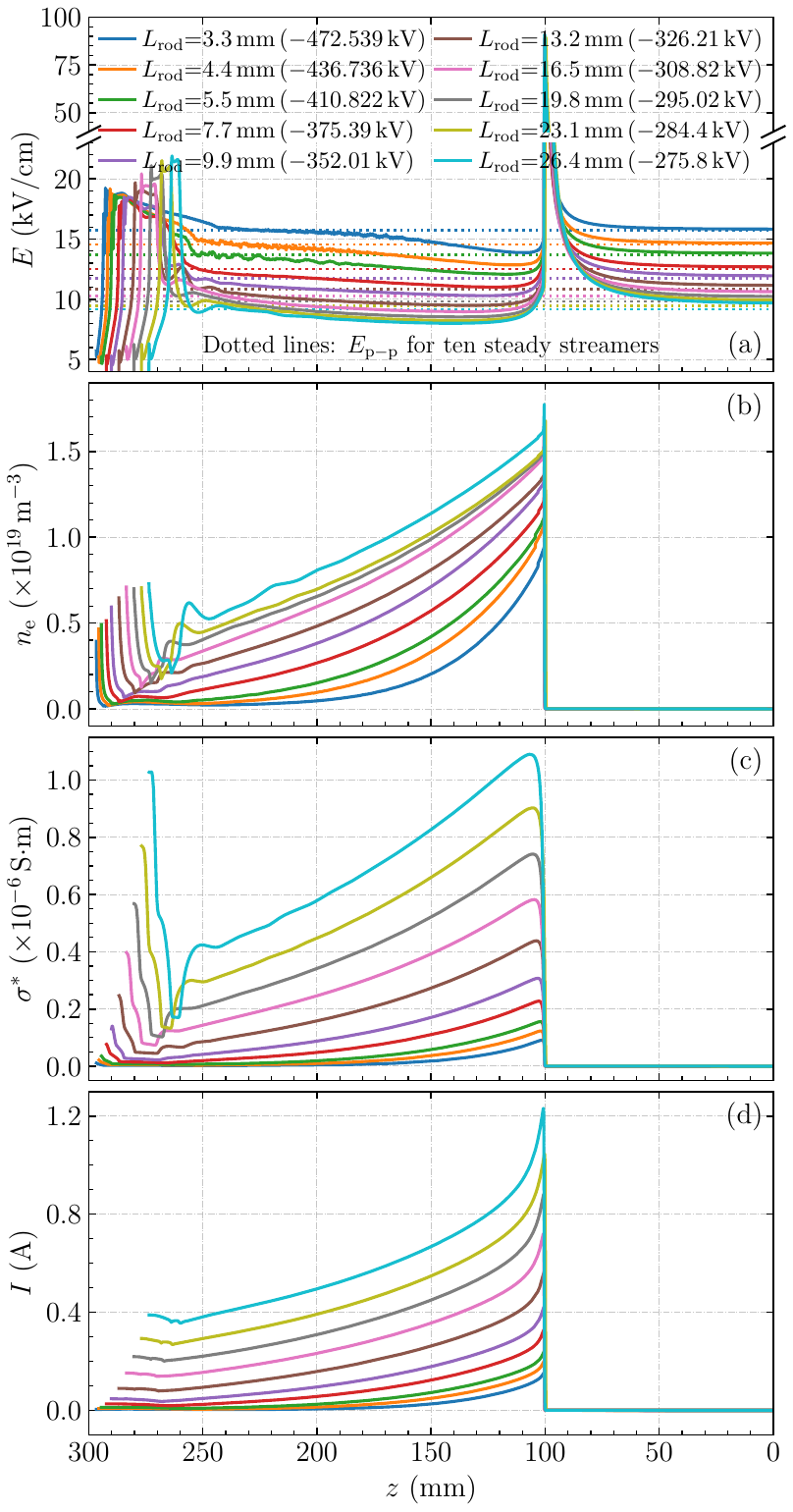}
    \caption{(a) The on-axis electric field $E$ and the background electric field $E_{\mathrm{p-p}}$, (b) the on-axis electron density $n_{\mathrm e}$, (c) the line conductivity $\sigma^*$ and (d) the electron conduction current $I$ for the ten steady streamers at $z_{\mathrm{head}}=100$\,mm corresponding to \fref{fig:ne for ten steady streamers}.
    All streamers propagate towards the right.}
    \label{fig:axial profiles of ten steady streamers}
\end{figure}

Axial profiles corresponding to these steady streamers at $z_{\mathrm{head}}=100$\,mm are also shown in \fref{fig:axial profiles of ten steady streamers}.
Here the line conductivity $\sigma^*$ was integrated radially as
\begin{equation}
\label{eq:line conductivity}
  \sigma^*(z) = 2 \pi e \int_0^{r_\mathrm{max}} r n_\mathrm{e} \mu_\mathrm{e} \, \mathrm{d} r\,,
\end{equation}
where $e$ is the elementary charge and $r_\mathrm{max}$ is a few times larger than the streamer radius.

For these ten rod electrodes, the background electric field $E_{\mathrm{p-p}}$ ranges from 9.19\,kV/cm to 15.75\,kV/cm for $L_\mathrm{rod}$ from 26.4\,mm to 3.3\,mm.
Typical values range from 10\,kV/cm to 12.5\,kV/cm, but for short electrodes the required $E_{\mathrm{p-p}}$ rapidly increases.
We could not simulate steady streamers for longer or shorter rod electrodes in domain B, due to branching for longer electrodes and due to the difficulty in locating the steady regime for shorter electrodes.

As was discussed in \sref{sec:stab-fields-steady}, $E_{\mathrm{p-p}}$ is closely related to the steady propagation field.
Our results therefore confirm the conclusion in~\cite{qin2014} that steady propagation fields can lie in a wide range, depending not only on the gas but also on the streamer's properties.
Since streamer properties are determined by many factors, including e.g., the electrode geometry, the applied voltage waveform and initial conditions, this explains part of the spread in experimentally measured stability fields~\cite{phelps1976, griffiths1976a, allen1991, niemeyer1995, allen1995, allen1999}.

That a longer electrode requires a lower $E_{\mathrm{p-p}}$ for steady propagation is in agreement with the results for steady positive streamers in~\cite{li2022}.
In~\cite{li2022}, it was argued that the length over which a streamer has a significant conductivity depends on the product $v \tau$, where $v$ is the streamer velocity and $\tau$ a characteristic time scale for the loss of conductivity due to e.g., attachment.
With a longer electrode, a faster streamer emerges, which will therefore have a longer conductive length, so that a lower background field is sufficient for its field enhancement, as can be seen in \fref{fig:ne for ten steady streamers} and \ref{fig:axial profiles of ten steady streamers}.

\begin{figure}
    \centering
    \includegraphics[width=0.49\textwidth]{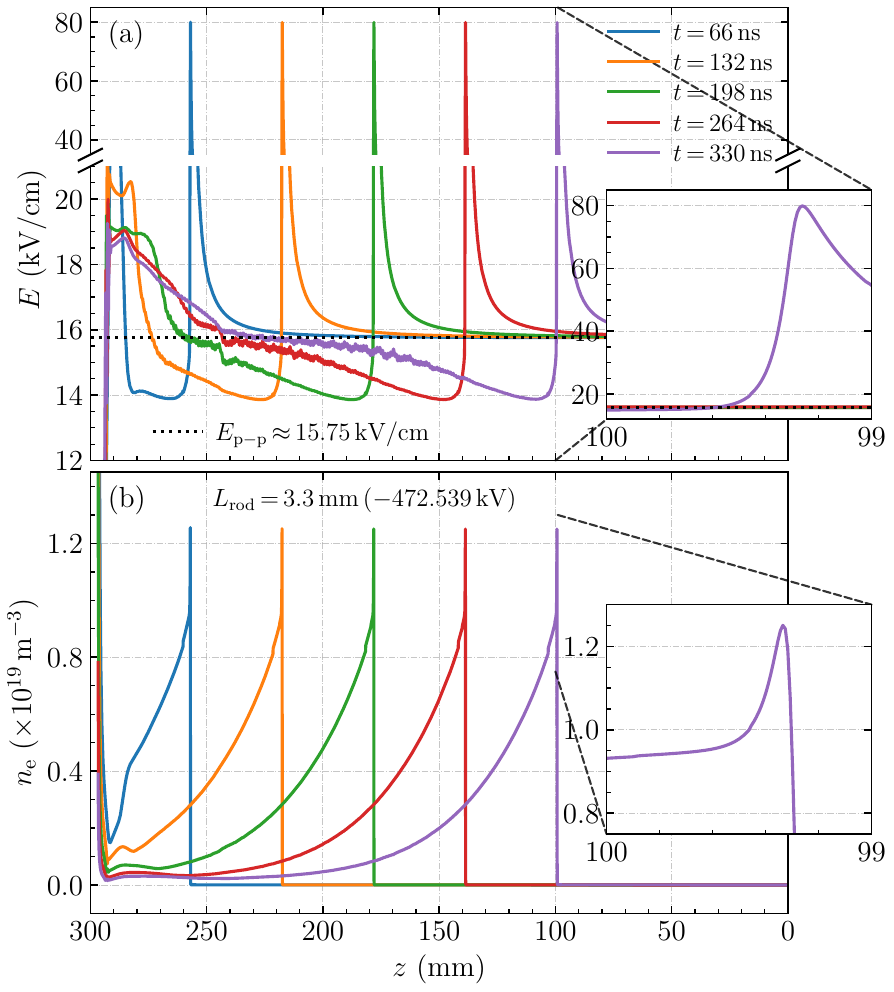}
    \caption{Time evolution of (a) the electric field $E$ and (b) the electron density $n_{\mathrm e}$ along $z$ axis for the steady streamer with $L_{\mathrm{rod}}=3.3$\,mm from \fref{fig:ne for ten steady streamers}.
    The background field $E_{\mathrm{p-p}} \approx$ 15.75\,kV/cm is shown in panel (a) for comparison.
    Zoomed views of $E$ and $n_{\mathrm e}$ at $t=330$\,ns and 99\,mm\,$\leqslant z \leqslant$\,100\,mm are included, revealing an overshoot in $n_{\mathrm e}$ in the streamer head.}
    \label{fig:E and ne for Lrod=3.3mm}
\end{figure}

\Fref{fig:E and ne for Lrod=3.3mm} shows the time evolution of axial profiles for the steady streamer with $L_\mathrm{rod} =$ 3.3\,mm.
As can be seen in \fref{fig:ne for ten steady streamers} and \ref{fig:E and ne for Lrod=3.3mm}, the conductive length is short enough to cause the internal field in the back of the streamer channel to return to its background field of 15.75\,kV/cm.
Like for steady positive streamers~\cite{francisco2021e, li2022}, steady negative streamers do not require a conductive channel to sustain the propagation, as will be discussed in \sref{sec:conductive channel}.
The electric field in the streamer channel ranges from about 14\,kV/cm to $E_{\mathrm{p-p}}$ and the attachment time ranges from 35--50\,ns, which coincidentally agrees well with the electron loss times for steady positive streamers in air~\cite{francisco2021e, li2022}.
However, the streamer velocity of about $6 \times 10^5$\,m/s in \fref{fig:E and ne for Lrod=3.3mm} is much larger than that of steady positive streamers (about 0.3--1.2\,$\times 10^5$\,m/s), which leads to a longer conductive length.
Another difference is that the electron density contains an overshoot near the head of negative streamers, as can be seen in \fref{fig:three examples profiles}(c1)--(c3), \ref{fig:axial profiles of ten steady streamers}(b) and \ref{fig:E and ne for Lrod=3.3mm}(b).

\begin{figure}
    \centering
    \includegraphics[width=0.49\textwidth]{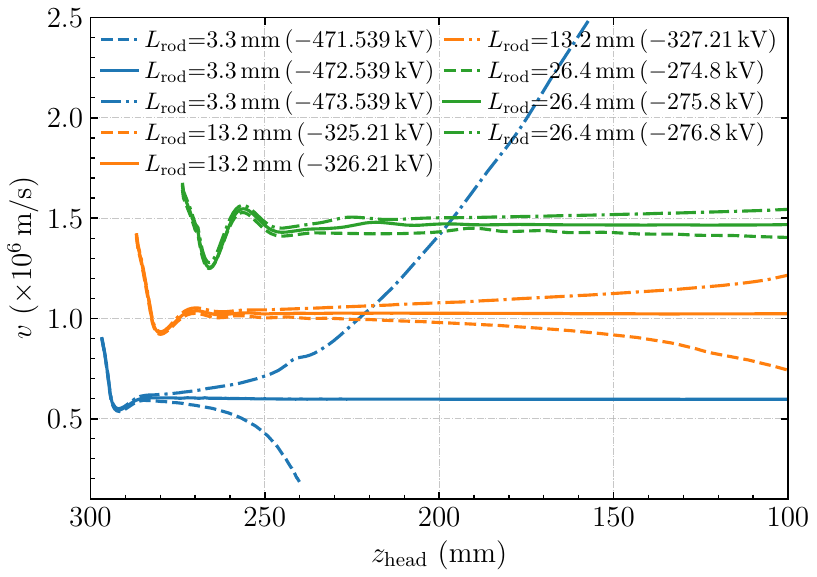}
    \caption{Illustration of the instability of steady negative streamers.
    The streamer velocity $v$ is shown versus the streamer position $z_{\mathrm{head}}$ for steady streamers with $L_{\mathrm{rod}}=$ 3.3\,mm, 13.2\,mm and 26.4\,mm, together with cases in which the applied voltages differ by $\pm$1\,kV from the respective steady values.
    % The simulations were performed in domain B until $z_\mathrm{head}=100$\,mm or fading, see \tref{tab:computational domain}.
    All streamers propagate towards the right.}
    \label{fig:v for EL=3-12-24mm}
\end{figure}

We remark that the steady streamers are unstable, in the sense that a tiny change in the applied voltage causes them to either accelerate or decelerate.
This instability is illustrated in \fref{fig:v for EL=3-12-24mm}, which shows streamer velocity versus position for cases with applied voltages that differ by $\pm$1\,kV from the steady values.
With these different applied voltages, the streamer continually accelerates or decelerates.
Hence, the steady streamer is not really stable, but exists only at the unstable boundary between accelerating and fading streamers.
Any minor change in input data or numerical parameters would also lead the steady streamer to eventually accelerate or decelerate.
The results in \fref{fig:v for EL=3-12-24mm} indicate that this instability is stronger for a slower negative steady streamer.

\section{Discussion and analysis}\label{sec:discussion and analysis}

Below, we discuss several important questions about negative streamers in air, and compare their properties with those of positive streamers.

% Maybe include at some point?
% In figure \ref{fig:three examples profiles} the field in the streamer channel $E_\mathrm{ch}$ is close to the `stability field'.
% For steady propagation, this has to be the case: if the field in the channel would be lower or higher, the head potential would respectively increase or decrease over time, so there would not be steady propagation.

\subsection{Does a negative streamer require a conductive channel to sustain its propagation?}\label{sec:conductive channel}

A key difference between negative and positive streamers is that negative streamers propagate in the same direction as the electron drift velocity, whereas positive streamers propagate opposite to it.
This means that positive streamers `suck up' free electrons ahead of them, a bit like a vacuum cleaner, which come out at the back of the streamers.
In contrast, a negative streamer `emits' electrons in the forward direction, figuratively speaking like a garden hose.
The excess of these electrons (compared to the positive ion density) forms the negative charge layer that provides field enhancement.

In case of steady propagation, which is an unstable mode, the net charge in the streamer head region is conserved.
The electric field, modified by the streamer, is then such that the net charge simply translates, while ionization is created at the streamer tip and lost at its back.
A negative streamer therefore does not require a conductive channel in order to sustain its propagation, as shown in \fref{fig:ne for ten steady streamers}.
The same observation was recently made for positive streamers in~\cite{francisco2021e}, which were referred to as `solitary streamers'.

Furthermore, our results show that an accelerating negative streamer also does not require a conductive channel connected to an electrode.
In such streamers, the head charge increases over time, but this head charge can be supplied by polarizing the channel behind the streamer.

% If it instead decelerates, negative charge is deposited behind the streamer head, reducing the head charge.

\subsection{Is there a minimal steady negative streamer?}\label{sec:minimal streamer}

For positive streamers, the concept of a minimal streamer, with a certain minimal diameter $d_\mathrm{min}$, was postulated in \cite{briels2006}.
In the experiments of~\cite{briels2008a}, a relation $p \cdot d_\mathrm{min}= 0.20\,\pm\,0.02$\,mm\,bar for the minimal optical diameter was found in air at room temperature, at pressures from 0.013 to 1 bar.
In nitrogen, streamers were found to be thinner, with $p \cdot d_\mathrm{min}= 0.12\,\pm\,0.03$\,mm\,bar.
By improving gas purity and optical diagnostics, similar trends but with somewhat lower values of $p \cdot d_\mathrm{min}$, namely 0.12\,mm\,bar for air and 0.07\,mm\,bar for nitrogen, were found in~\cite{nijdam2010}.
Furthermore, in the theoretical and numerical study of~\cite{naidis2009}, the minimal optical diameters of positive streamers in air at atmospheric pressure were related to $E_\mathrm{max}$.
For $E_\mathrm{max}=$ 140\,kV/cm and 160\,kV/cm, minimal diameters of 0.27\,mm and 0.20\,mm were respectively predicted, in the limit of zero streamer velocity.
In~\cite{qin2014}, steady propagation for both positive and negative streamers in air at atmospheric pressure was simulated.
Smaller minimal diameters of 0.072\,mm for positive streamers and of about 0.5\,mm for negative streamers were suggested, where the diameter was defined as the FWHM of the electron density.

In another numerical study of~\cite{starikovskiy2020}, a minimal eletrodynamic diameter ($d_{Er}$, see \ref{sec:diameter comparison}) of about 1.3\,mm in air at atmospheric pressure was observed for negative streamers.
Experimentally, few measurements appear to be available for minimal negative streamer diameters.
In~\cite{briels2008}, the negative streamer with a minimal optical diameter of 1.2\,mm was observed in air at 1 bar.
However, negative streamer inception required a rather high applied voltage in the above two studies, and thinner negative streamers could be generated with a different electrode geometry and applied voltage waveform.

\begin{figure}
  \centering
  \includegraphics[width=0.48\textwidth]{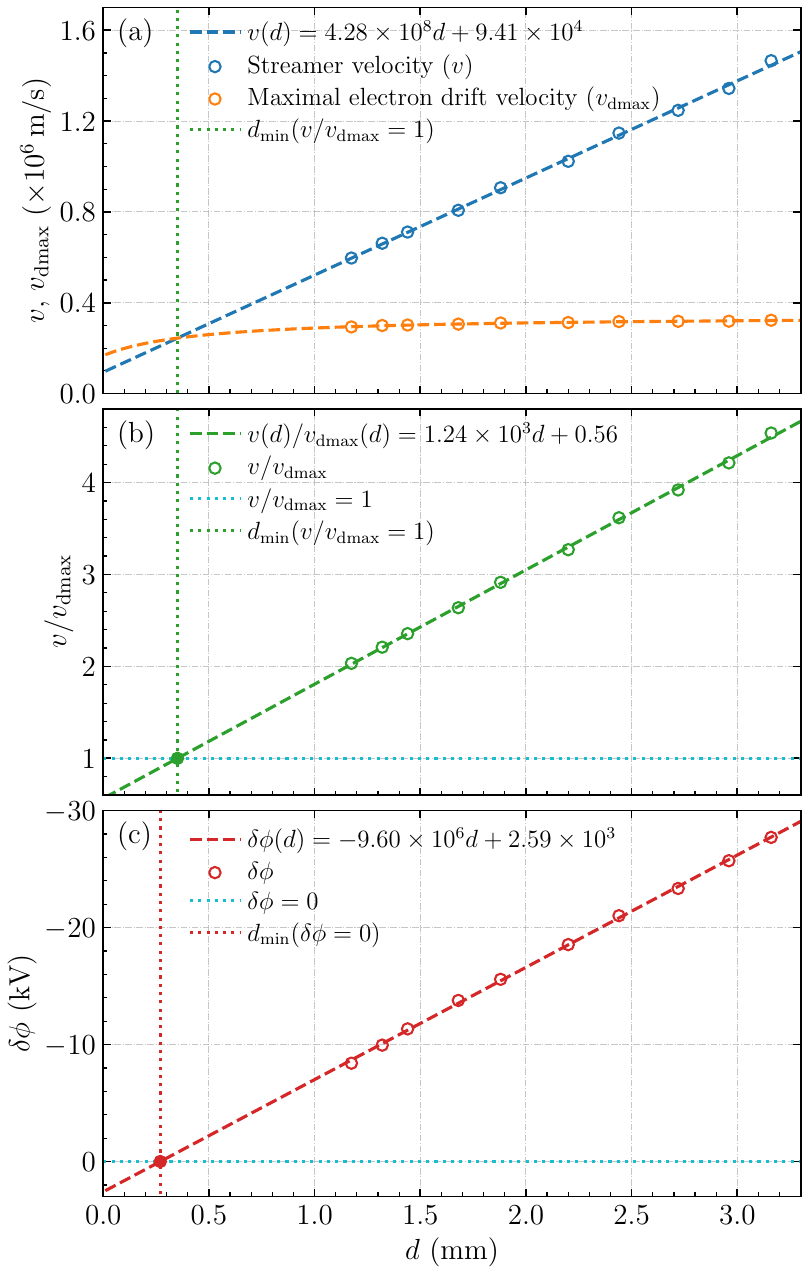}
  \caption{Linear relationships between the optical streamer diameter $d$ and (a) the streamer velocity $v$, (b) the ratio $v/v_\mathrm{dmax}$ and (c) the head potential $\delta \phi$, respectively.
    The maximal electron drift velocity $v_{\mathrm{dmax}}$ is also shown in panel (a).
    The open circles \opencircle correspond to the steady cases shown in figure \ref{fig:ne for ten steady streamers}.
    Estimates of a minimal optical diameter $d_{\mathrm{min}}$ for the steady streamers are indicated by full circles \fullcircle.
    These estimates are given by: (b) the diameter at which $v / v_{\mathrm{dmax}} = $ 1 and (c) the diameter at which $\delta \phi = $ 0.}
  \label{fig:dmin-optical}
\end{figure}

From our simulations, we can estimate a lower bound for the minimal optical diameter of steady negative streamers in air.
The variation of their optical diameter $d$ in different background fields is shown in \fref{fig:quantities versus d}.
The streamer optical diameter has approximately linear relationships with several quantities: the streamer velocity $v$, % the quantity $E_{\mathrm{p-p}} \times d$,
the ratio between streamer velocity and maximal electron drift velocity $v/v_\mathrm{dmax}$, and the head potential $\delta \phi$, as shown in \fref{fig:dmin-optical}.
These linear relationships can be described by
\begin{equation}
  \label{eq:v versus d}
  v(d) = 4.28 \times 10^8d + 9.41 \times 10^4\,,
\end{equation}
\begin{equation}
  \label{eq:velocity ratio versus d}
  v(d)/v_\mathrm{dmax}(d) = 1.24 \times 10^3d + 0.56\,,
\end{equation}
\begin{equation}
  \label{eq:dphi versus d}
  \delta \phi(d) = -9.60 \times 10^6d + 2.59 \times 10^3\,,
\end{equation}
% \begin{equation}
%   \label{eq:Ebg*d versus d}
%   E_{\mathrm{p-p}}(d) \times d = 5.39 \times 10^5d + 1.20 \times 10^3\,,
% \end{equation}
where all above quantities have for simplicity of notation been divided by their respective units (using meters, seconds and volts).
Based on these relations, rough estimates can be made for a lower bound on the minimal optical diameter $d_\mathrm{min}$:
\begin{itemize}
  \item The diameter at which $v = v_\mathrm{dmax}$\,, which gives $d_\mathrm{min} = 0.35$\,mm.
  \item The diameter at which $\delta \phi = 0$\,, which gives $d_\mathrm{min} = 0.27$\,mm.
  Note that in this theoretical limit there is essentially no space charge, and thus no light emission to deduce an optical diameter from.
  % \item The diameter at which $E_{\mathrm{p-p}} = E_\mathrm{k}$ (28\,kV/cm), which gives $d_\mathrm{min} = 0.53$\,mm. \ue{\it Why this criterion?}
\end{itemize}
So for the minimal optical diameter, these rough lower bounds give 0.27\,mm to 0.35\,mm, smaller but comparable to the measured minimal optical diameter of 1.2\,mm found in~\cite{briels2008}.
% We remark that $d_\mathrm{min}$ is probably larger than these estimates, since the criteria $\delta \phi = 0$ and $v = v_\mathrm{dmax}$ give a rather crude lower bound, as will be further discussed in \sref{sec:definition of fading}.

\subsection{The ratio between streamer velocity and electron drift velocity}\label{sec:velocity ratio}

% \ue{A FEW BASIC REMARKS ON THIS SECTION:\\
% For negative streamers without diffusion, photoionization or background ionization, we know that the streamer velocity is precisely equal to the electron drift velocity in the maximal electric field. There is a shock front of electron density, and behind it $\partial_z n_e >0$. (Not $<0$ as used in the argument below.) This is, of course a spatial case, but included in our equations. I can guide you to some of my analytical work published between 1997 and 2004 where I elaborate this.\\
% Furthermore in the equation below, $\partial_z v_d$ should be removed, as I already discuss below, but there is also the nonlinear term $\mu n_e \nabla\cdot{\bf E}$ in the front region that comes from to $\nabla\cdot \left(\mu {\bf E} n_e\right)$ and that is not included in the argument below.\\
% I suggest that I would shorten and rewrite this section. But we first should agree.
% \\
% }

For negative streamers, the ratio between streamer velocity and maximal electron drift velocity $v/v_\mathrm{dmax}$ should be greater than one~\cite{ebert1997, nijdam2020a,starikovskiy2021}.
In our simulations, $v/v_\mathrm{dmax}$ ranges from about 2.0 to 4.5 for steady negative streamers.
In contrast, this ratio was observed to be about 0.05 to 0.26 for steady positive streamers in~\cite{li2022}.
% \ue{\it This argument needs rewriting. I can do it, after we have discussed the issue.
% Naidis' equation below is wrong, because he naively replaces $\nabla\cdot {\bf E}$ by $\partial_z E$. But $\nabla\cdot {\bf E}=0$ in the active zone where space charge is negligible! In my papers in PRL 1996 and PRE 1997, I have already discussed the polarity dependence of $v\pm v_d$ for planar fronts (see section IV in the 1997 paper, where $v\pm v_d$ is denoted as $v\pm E$). Of course, these fronts are planar.}
To understand why these ratios are so different for steady negative and positive streamers, it is useful to consider an analytic approximation for uniformly translating streamers, as was done in~\cite{naidis2009}.
If we assume that $\nabla \cdot (\mu_{\mathrm e} \boldsymbol{\mathrm E} n_{\mathrm e}) \approx \mu_{\mathrm e} \boldsymbol{\mathrm E} \cdot \nabla n_{\mathrm e}$ (i.e., $\mu_{\mathrm e}$ is assumed constant and $\nabla \cdot \boldsymbol{\mathrm E} = 0$), and if we ignore effects due to diffusion and photoionization, we can derive the following expression
\begin{equation}
  \label{eq:uniformly-translating}
  -\frac{v}{v_\mathrm d} \frac{\partial_z n_\mathrm e}{n_\mathrm e}
  \pm \frac{\partial_z n_\mathrm e}{n_\mathrm e} = \bar{\alpha}\,,
\end{equation}
where $v$ is the streamer velocity (in the $+z$ direction), $v_\mathrm d = |\mu_\mathrm e E|$ the absolute value of the electron drift velocity, $\bar{\alpha}$ the effective ionization coefficient, and the $+$ sign corresponds to negative streamers and the $-$ sign to positive ones. 
% \ue{\it The term $\partial_z v_{\mathrm d}$ should be absent, if $\mu_e$ is constant. And for which value of $z$ do you evaluate the equation? This is a rhetoric question --- let's discuss first and then rewrite.}
If we introduce the length scale $\lambda_{n_\mathrm e} = n_\mathrm e / |\partial_z n_\mathrm e|$ and consider the fact that $\partial_z n_\mathrm e$ is negative for a streamer propagating in the $+z$ direction, equation \eref{eq:uniformly-translating} can be written as
\begin{eqnarray}
  \label{eq:uniformly-translating-simple}
  |v|/v_\mathrm d = \bar{\alpha} \lambda_{n_\mathrm e} \pm 1\,,
\end{eqnarray}
with the $+$ sign corresponding to negative streamers and the $-$ sign to positive ones.
Note that $v$ is assumed constant here, whereas $v_\mathrm d$, $\bar{\alpha}$ and $\lambda_{n_\mathrm e}$ all vary ahead of a streamer.

Even without knowing the values of $\bar{\alpha}$ and $\lambda_{n_\mathrm e}$, a couple of statements can be made about the ratio $v/v_\mathrm{dmax}$.
First, for steady positive streamers, $v/v_\mathrm{dmax}$ can in principle be arbitrarily small, in agreement with the value of $0.05$ mentioned above.
The limit of zero streamer velocity corresponds to $\lambda_{n_\mathrm e} = 1/\bar{\alpha}$, and thus a density profile proportional to exp$\left(- \int \bar{\alpha}\,\mathrm{d} z \right)$.
On the other hand, for steady negative streamers, $v/v_\mathrm{dmax}$ cannot be lower than one, as expected.
In high background fields streamers usually accelerate, but if their properties do not change rapidly, equation \eref{eq:uniformly-translating-simple} is still approximately valid.
The fact that negative and positive streamers become similar in high background fields~\cite{babaeva1997, briels2008, starikovskiy2020} therefore suggests that the term $\bar{\alpha} \lambda_{n_\mathrm e}$ then becomes large.

If we hypothetically assume that a steady positive and negative streamer exist with the same values of $\bar{\alpha}$ and $\lambda_{n_\mathrm e}$, then equation \eref{eq:uniformly-translating-simple} implies that $v/v_\mathrm{dmax} \geqslant 2$ for the steady negative streamer, since this ratio cannot be negative for the steady positive streamer.
However, $\bar{\alpha}$ and $\lambda_{n_\mathrm e}$ are generally different for positive and negative streamers.
In simulations in another computational domain, which are not presented here, we have actually found steady negative streamers for which $v/v_\mathrm{dmax}$ could be as low as $1.8$.
Note that without photoionization and background ionization, the ratio $v/v_\mathrm{dmax}$ is expected to be close to one~\cite{ebert1996}.

\subsection{The definition of negative streamer fading}\label{sec:definition of fading}

% \ue{\it I think that the essentials of the fading are:
% the field enhancement is getting so low, that the field approaches the breakdown field $E_k$, and therefore the ionization is essentially stopping, so light emission is very weak as well. Then the electrons just drift, and you naturally get $v=v_d$. Because the existing electrons drift radially outwards, the field enhancement decreases further, and you get a continuous streamer-to-avalanche transition. Is there a real transition point or is the transition continuous? Light emission does not stop for fields below $E_k$, but there is a morphological change. Should we base our criterion on this morphology, or on $E_k$? A clearly defined morphological criterion could be the loss of the electron overshoot in the front, as negative streamers always have a maximum of $n_e$ in the space charge layer.}

When the background electric field is too low to sustain steady propagation, a negative streamer fades out and loses its field enhancement, and the streamer velocity becomes comparable to the maximal electron drift velocity.
However, the discharge does not fully stop at some point, because the electrons that make up its space charge layer continue their drift motion, further lowering the velocity and field enhancement.
This makes it hard to uniquely define when a negative streamer has faded.
For the results presented in sections \ref{sec:three examples} and \ref{sec:fading streamers}, we have somewhat arbitrarily used the criterion $E_\mathrm{max} \leqslant 35$ kV/cm.
A related criterion could for example be
$$E_\mathrm{max}/\alpha(E_\mathrm{max}) \geqslant V_\mathrm{c}\,,$$
where $V_\mathrm{c}$ is for example 250\,V.
% \ue{Or $\bar{\alpha}(E_\mathrm{max}) = 0$.}
Note that $e E/\alpha(E)$, where $e$ is the elementary charge, can be interpreted as the energy an electron has to gain from the field per ionization.

We can also consider the ratio between streamer velocity and maximal electron drift velocity $v / v_\mathrm{dmax}$.
As was discussed in \sref{sec:velocity ratio}, a negative streamer needs to propagate with at least the maximal electron drift velocity.
Our results in \fref{fig:ten fading streamers} suggest that a criterion like $v / v_\mathrm{dmax} \leqslant \kappa$, with for example $\kappa = 1.5$, could be used to define the moment of negative streamer fading.

However, it was argued in~\cite{qin2014} that for steady negative streamers $E_\mathrm{max}$ can be as weak as $\sim$\,1.2$E_\mathrm{k}$, and that the ratio $v / v_\mathrm{dmax}$ can be as low as $\sim$\,1.03.
If that is the case, it is hard to distinguish between a fading negative streamer and an electron avalanche with weak field enhancement.

Finally, we remark that negative streamer fading differs significantly from positive streamer stagnation.
When positive streamers stagnate, their field enhancement tends to increase~\cite{niknezhad2021b, li2022}.
Only after their velocity has become comparable to the ion drift velocity, which is much lower than the electron drift velocity, do they lose their field enhancement.

\subsection{Comparison between steady negative and positive streamers}\label{sec:steady-neg-and-pos}

We can quantitatively compare our simulated steady negative streamers with the results for steady positive streamers in~\cite{li2022}.
We consider the steady negative and positive cases corresponding to the lowest and highest respective background electric fields, and several properties of these streamers are compared in \tref{tab:steady-comparison}.
Velocity, radius, ratio between streamer velocity and maximal electron drift velocity and the maximal electron conduction current all differ by more than an order of magnitude between these steady negative and positive streamers.
% Due to its higher velocity, the conductive length of the steady negative streamer is significantly longer than that of the positive steady one.
Due to their higher velocities, the conductive lengths of the two steady negative streamers are respectively significantly longer than that of the positive steady ones.
% \sout{The largest difference is in the maximal electron conduction current, which is approximately two orders of magnitude larger for the negative steady streamer.}
% \sout{In addition, the maximal electron density, maximal positive ions density, maximal line charge density and minimal internal on-axis electric field all differ by a factor of about seven.}

Given these major differences, it is perhaps a bit surprising that the steady propagation fields for these negative and positive cases differ by a much smaller factor. 
% only by a factor of about 2.5.

\begin{table*}
\centering
\caption{\label{tab:steady-comparison}Comparison between steady negative and positive streamers, corresponding to the lowest and highest respective background electric fields.
For the steady negative streamers, properties were obtained at the moment corresponding to \fref{fig:axial profiles of ten steady streamers}, and for the positive ones they were obtained from~\cite{li2022}.}
\lineup
\begin{tabular*}{0.98\textwidth}{l @{\extracolsep{\fill}} | l l l | l l l}
  \br
  & \multicolumn{3}{c|}{Lowest background field} & \multicolumn{3}{c}{Highest background field} \\
  Quantity & Negative & Positive & Ratio & Negative & Positive & Ratio\\
  \mr
  $v$ (m/s) & $1.47 \times 10^6$ & $1.2 \times 10^5$ & 12.25 & $5.97 \times 10^5$ & $3 \times 10^4$ & 19.90 \\
  $R_{Er}$ (mm)$^{\mathrm a}$ & 3.15 & 0.11 & 28.64 & 1.25 & 0.036 & 34.72\\
  $E_{\mathrm{p-p}}$ (kV/cm) & 9.19 & 4.05 & 2.27 & 15.75 & 5.42 & 2.91\\
  $E_\mathrm{max}$ (kV/cm) & 91 & 152 & 0.60 & 80 & 222 & 0.36\\
  $E_\mathrm{min}$ (kV/cm)$^{\mathrm b}$ & 8.02 & 1.17 & 6.85 & 13.87 & 0.42 & 33.02\\
  $\max(n_\mathrm e)$ (m$^{-3}$) & $1.77 \times 10^{19}$ & $1.11 \times 10^{20}$ & 0.16 & $1.25 \times 10^{19}$ & $3.12 \times 10^{20}$ & 0.04\\
  $\max(n_{\mathrm i}^+)$ (m$^{-3}$)$^{\mathrm c}$ & $1.68 \times 10^{19}$ & $1.13 \times 10^{20}$ & 0.15 & $1.00 \times 10^{19}$ & $3.30 \times 10^{20}$ & 0.03\\
  $I_\mathrm{max}$ (A) & 1.23 & $1.38 \times 10^{-2}$ & 89.13 & 0.16 & $1.91 \times 10^{-3}$ & 83.77\\
  $\lambda_\mathrm{max}$ (C/m) & \-$8.41 \times 10^{-7}$ & $1.11 \times 10^{-7}$ & \-7.58 & \-$2.71 \times 10^{-7}$ & $5.44 \times 10^{-8}$ & \-4.98\\
  $\sigma^*_\mathrm{max}$ (S$\cdot$m) & $1.09 \times 10^{-6}$ & $5.09 \times 10^{-8}$ & 21.41 & $9.13 \times 10^{-8}$ & $2.38 \times 10^{-8}$  & 3.84\\
  $v_\mathrm{dmax}$ (m/s) & $3.23 \times 10^5$ & $4.7 \times 10^5$ & 0.69 & $2.94 \times 10^5$ & $6.1 \times 10^5$ & 0.48\\
  $v / v_\mathrm{dmax}$ & 4.55 & 0.26 & 17.50 & 2.03 & 0.049 & 41.43\\
  $\delta \phi$ (kV) & \-27.61 & 2.83 & \-9.42 & \-8.37 & 1.52 & \-5.51\\
  % \mr
  \br
\end{tabular*}
\small
\begin{tabular*}{0.98\textwidth}{@{\extracolsep{\fill}}l}
$^{\mathrm a}$ $R_{Er}$ is the electrodynamic radius at which the radial electric field has a maximum, see \ref{sec:diameter comparison}.\\
$^{\mathrm b}$ $E_\mathrm{min}$ is the minimal internal on-axis electric field behind the streamer head.\\
$^{\mathrm c}$ $n_{\mathrm i}^+$ is the total number density of all positive ion species.\\
\end{tabular*}
\end{table*}

\paragraph{\textbf{Estimation of the internal field.}}

\begin{figure}
  \centering
  \includegraphics[width=0.48\textwidth]{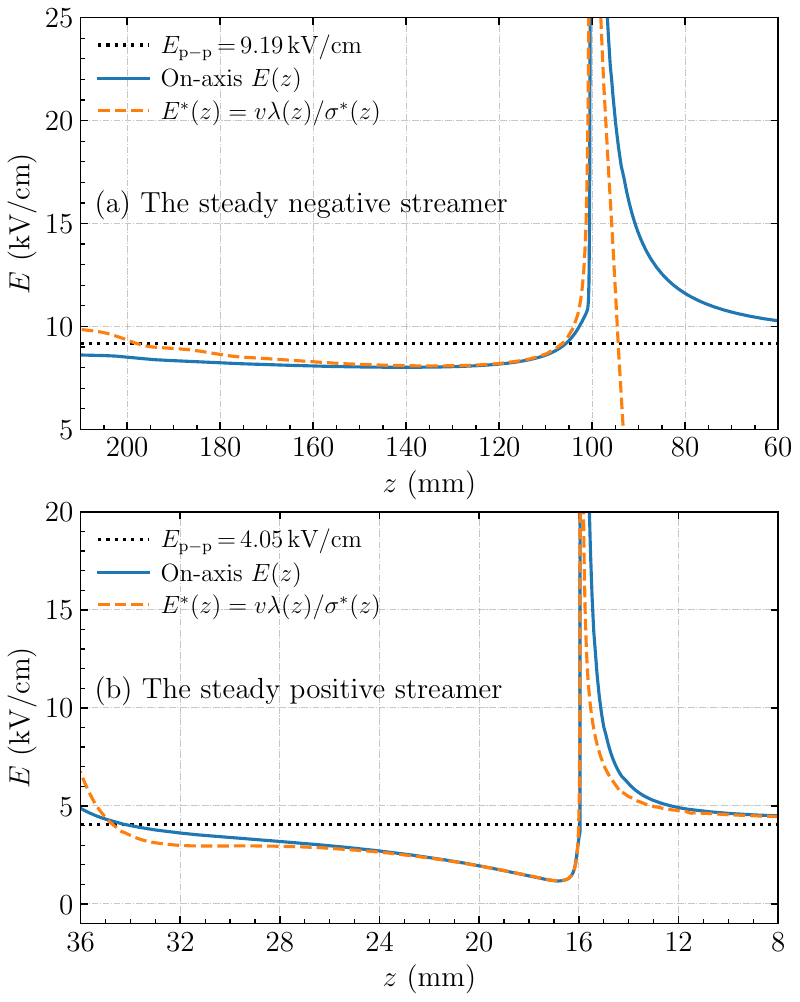}
  \caption{The average electric field $E^*(z)$ in the $z$ direction according to equation \eref{eq:internal-field}, together with the actual on-axis electric field $E(z)$ for steady negative and positive streamers from table \ref{tab:steady-comparison} corresponding to the lowest background fields.
  Both streamers propagate towards the right.}
  \label{fig:internal field}
\end{figure}

We now compare the steady negative and positive streamers from the point of view of their electron conduction current, line charge density, line conductivity and internal electric field.
If a streamer translates approximately uniformly with velocity $v$, its line charge density $\lambda$ should translate approximately uniformly with the same velocity.
The evolution of $\lambda$ is then given by $\partial_t \lambda = -\partial_x I = -v \partial_x \lambda$, where $I$ is the electron conduction current and $v$ is the streamer velocity.
General solutions are the form $I(z) = v \lambda(z) + I_0$, but since there is no current ahead of the streamer, $I_0 = 0$, and
% After integrating, the result is $I = v \lambda + I_0$ \ue{with $d_zI_0=0$}, and since there is no current \ue{ahead of} the streamer, we have \ue{$I_0 \equiv 0$ and}
\begin{equation}
  \label{eq:current-line-charge}
  I(z) = v \lambda(z)\,.
\end{equation}
On the other hand, we can also write the current as
\begin{equation}
  \label{eq:current-internal}
  I(z) = \sigma^*(z) E^*(z)\,,
\end{equation}
where $\sigma^*(z)$ is the streamer's line conductivity given by equation \eref{eq:line conductivity} and $E^*(z)$ is the average electric field in the $z$ direction.
If equations \eref{eq:current-line-charge} and \eref{eq:current-internal} are combined, the result is
\begin{equation}
  \label{eq:internal-field}
  E^*(z) = v \lambda(z) / \sigma^*(z)\,.
\end{equation}
\Fref{fig:internal field} shows the actual on-axis electric field for the positive and negative steady streamers from \tref{tab:steady-comparison} corresponding to the lowest background fields, together with results from equation \eref{eq:internal-field}.
Behind the streamer head, $E^*(z)$ agrees well with the on-axis electric field $E(z)$ for both steady streamers.
Note that discrepancies are visible ahead of the streamer, where equation (\ref{eq:internal-field}) does not apply.
% There are also small discrepancies just behind the head, where the on-axis electric field there differs from the average electric field in the $z$ direction.

The minimal internal on-axis electric fields $E_\mathrm{min}$ behind the streamer head are 8.02\,kV/cm and 1.17\,kV/cm for the above-mentioned steady negative and positive streamers, respectively.
These internal fields are related to the steady propagation fields, which they cannot exceed.
There is however a polarity difference.
For steady positive streamers, the internal field can be significantly below the background field \cite{li2022}, whereas for steady negative streamers these fields are rather similar.
In other words, the screening of the electric field is weaker for steady negative streamers.

Our observations of steady negative streamers can be related to equation \eref{eq:internal-field} as follows.
We find a lower steady propagation field for a faster steady streamer, which also has a higher line charge density $\lambda$.
Equation \eref{eq:internal-field} suggests that this can only be the case when the line conductivity $\sigma^*$ increases more rapidly than the product $v \lambda$, which is primarily driven by an increase in the radius.

\section{Conclusions and outlook}\label{sec:conclusions and outlook}

\subsection{Conclusions}

We have simulated single negative streamers in air at 300\,K and 1 bar in 125\,mm and 300\,mm long gaps, using a 2D axisymmetric fluid model.
With the same initial conditions, accelerating, steady and fading negative streamers were obtained by changing the background electric field.
The properties of the steady streamers remained nearly constant during their propagation.
However, this steady propagation mode is not stable, in the sense that a small change in parameters or streamer properties causes them to accelerate or decelerate.
Our main conclusions are listed below:
\begin{itemize}
  \item With different high-voltage electrode geometries, steady propagation for negative streamers was obtained in background fields ranging from 9.19\,kV/cm to 15.75\,kV/cm.
  The lowest background field was obtained for the longest electrode and the fastest steady streamer.
  This confirms that there is no unique steady propagation field (or stability field) for such streamers, as was previously suggested in~\cite{qin2014} and as was recently found for positive streamers~\cite{li2022}.
  \item Steady negative streamers are able to keep propagating over tens of centimeters with only a finite conductive length behind their heads, similar to steady positive streamers~\cite{francisco2021e, li2022}.
  They transport a constant amount of charge, and their propagation resembles a solitary wave.
  The conductivity in the back of the streamer channel disappears mainly due to attachment.
  \item The ratio between streamer velocity and maximal electron drift velocity $v/v_\mathrm{dmax}$ for steady negative streamers is much higher than that of steady positive streamers.
  For steady positive streamers, $v/v_\mathrm{dmax}$ can be as small as 0.05~\cite{li2022}.
  In contrast, for steady negative streamers, $v/v_\mathrm{dmax}$ cannot be lower than one, and the minimum ratio we observe is two.
  \item When we compare the steady negative and positive streamers corresponding to the lowest and highest respective background fields, their properties are totally different.
  The velocity, radius, $v/v_\mathrm{dmax}$,  and maximal electron conduction current all differ by more than an order of magnitude.
  In contrast, the lowest steady propagation fields of 9.19\,kV/cm (negative) and 4.05\,kV/cm (positive) differ only by a factor of about two.
%   In contrast, the steady propagation fields for negative and positive steady streamers differ only by a factor of about 2.5.
  \item For steady negative streamers, we observe approximately linear relationships between the optical diameter and the streamer velocity, the steamer head potential and the ratio $v/v_\mathrm{dmax}$.
  From these linear relations, rough estimates can be made for a lower bound on the minimal optical diameter, which lie between 0.27\,mm and 0.35\,mm.
\end{itemize}

\noindent In addition, we have the following minor conclusions:
\begin{itemize}
  \item In different background fields, negative streamer fading is highly similar, although fading occurs earlier in a lower background field.
  The channel conductivity disappears, and the ratio $v/v_\mathrm{dmax}$ decreases to about one. 
  \item It is hard to uniquely define when a negative streamer has faded.
  We have proposed several criteria, based on e.g., the maximal electric field or the ratio $v/v_\mathrm{dmax}$.
  \item There are different definitions of the streamer diameter.
  The electrodynamic diameter is usually larger than the optical diameter, with an approximately constant ratio of about two between them for steady negative streamers.
\end{itemize}

\subsection{Outlook}

An interesting question for future work would be how the steady propagation field depends on e.g., the applied voltage waveform and the initial conditions.
Understanding the relationships that were observed between several streamer properties requires further analysis as well.
Finally, future research could implement more realistic boundary conditions on the high-voltage electrode, for example, by including electron emission, to more realistically study the inception of negative streamers.

\ack
B.G. was funded by the China Scholarship Council (CSC) (Grant No.\,201906280436), and X.L. was supported by NWO-STW-project 15052 `Let CO$_2$ Spark!'.
We acknowledge fruitful discussions with Zhen Wang, Hani Francisco and Hemaditya Malla at CWI.
We thank both anonymous referees for their valuable suggestions.

\section*{Data availability statement}
% The source code and documentation for the model used in the present paper are available at \url{https://gitlab.com/MD-CWI-NL/afivo-streamer} (git commit \url{4ce20a0b}) and at \url{https://teunissen.net/afivo_streamer}.
The data that support the findings of this study are openly available at the following URL/DOI: \url{https://doi.org/10.5281/zenodo.6400856}.

\appendix

\section{Diameter comparison}\label{sec:diameter comparison}

In addition to the optical diameter $d_{\mathrm{optical}}$ defined in \sref{sec:definitions}, we here give two other definitions of the streamer diameter.
First, there is the electrodynamic diameter $d_{Ez} = 2 \times R_{Ez}$, based on the decay of the on-axis electric field ahead of the streamer head.
Near the head, this field decay is approximately quadratic, like that of a charged sphere, which depends on the streamer radius.
Ahead of the streamer ($z > z_\mathrm{head}$), it can be approximated by:
\begin{equation}
  \label{eq:E-fit}
  E_\mathrm{fit}(z) = E_\mathrm{bg} + (E_\mathrm{max} - E_\mathrm{bg}) \left(1 + \frac{z-z_\mathrm{head}}{R_{Ez}} \right)^{-2},
\end{equation}
where $E_\mathrm{bg}$ is the axial background electric field and $E_\mathrm{max}$ is the maximal electric field at the streamer head.
Since $E_\mathrm{bg}$ and $E_\mathrm{max}$ are known, we can obtain the radius $R_{Ez}$ by locating the $z$ coordinate at which the actual electric field is equal to $0.25 E_\mathrm{max} + 0.75 E_\mathrm{bg}$, which corresponds to $E_\mathrm{fit}(z_\mathrm{head} + R_{Ez})$.
Second, there is the electrodynamic diameter $d_{Er} = 2 \times R_{Er}$, where $R_{Er}$ is the radius at which the radial component of the electric field $E_r$ has a maximum~\cite{bagheri2020, francisco2021a, wang2022, li2022}.

\begin{figure}
    \centering
    \includegraphics[width=0.47\textwidth]{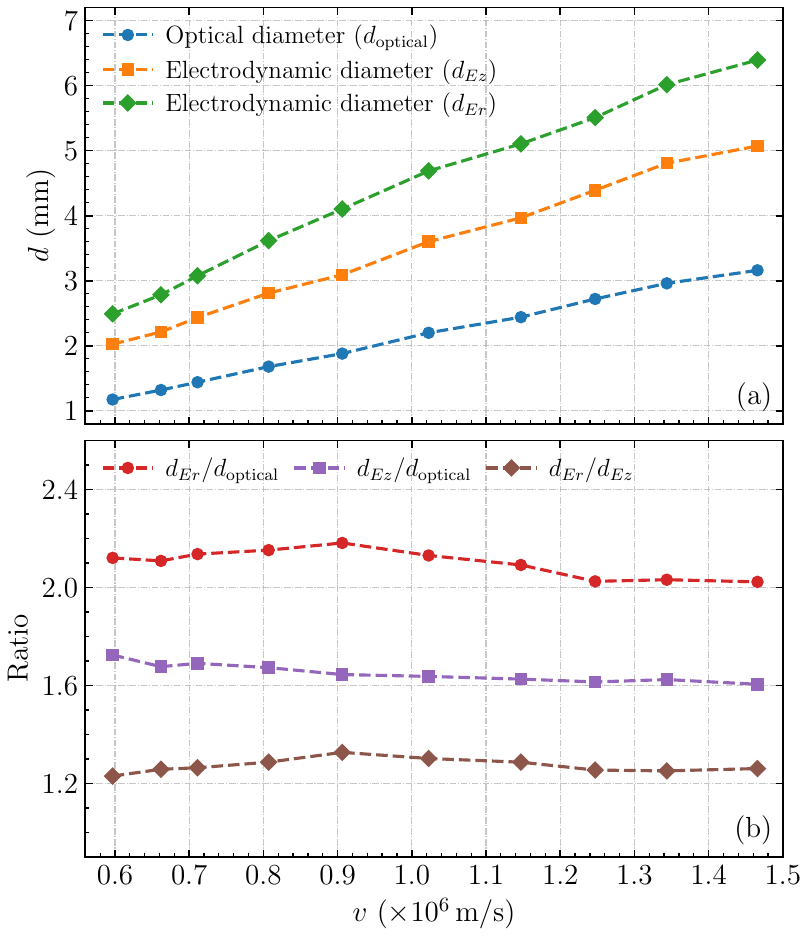}
    \caption{ (a) The optical diameter $d_{\mathrm{optical}}$, the electrodynamic diameter $d_{Ez}$ and the electrodynamic diameter $d_{Er}$ versus the streamer velocity $v$ for the ten steady streamers corresponding to \fref{fig:ne for ten steady streamers}.
    Each symbol represents one case, by taking an average over the propagation from $z_{\mathrm{head}}=$ 150\,mm to $z_{\mathrm{head}}=$ 100\,mm.
    (b) The ratios of $d_{Er} / d_{\mathrm{optical}}$, $d_{Ez} / d_{\mathrm{optical}}$ and $d_{Er} / d_{Ez}$ versus the streamer velocity $v$.}
    \label{fig:d comparison}
  \end{figure}

\Fref{fig:d comparison} shows these three definitions of the streamer diameter for the cases shown in figure \ref{fig:ne for ten steady streamers}, as well as the ratios between them.
% Approximately linear relationships between these three definitions of the streamer diameter and the streamer velocity are observed.
The ratios of $d_{Er} / d_{\mathrm{optical}}$, $d_{Ez} / d_{\mathrm{optical}}$ and $d_{Er} / d_{Ez}$ are approximately constant, and their values are about 2.1, 1.6 and 1.3, respectively.
The ratio $d_{Er} / d_{\mathrm{optical}} = 2.1$ agrees with the conclusion in~\cite{pancheshnyi2005c, nudnova2008, luque2009} that the electrodynamic diameter is about twice the optical diameter.

\section{Ion motion}\label{sec:ion motion}

Ion motion can be an important mechanism for slow positive streamers~\cite{niknezhad2021b, li2022}, especially when the streamer velocity becomes comparable to the ion drift velocity.
However, as was discussed in \sref{sec:velocity ratio}, a negative streamer needs to propagate with at least the maximal electron drift velocity, which is approximately two orders of magnitude larger than the ion drift velocity.
As a result, ion motion has little effect on the propagation of the negative streamers simulated in this paper.
Note that in regions where the ion density is much higher than the electron density ions contribute significantly to the plasma conductivity, but this conductivity is then much lower than it is near the streamer head.

\section{Effect of electron-ion recombination reactions}\label{sec:recombination time}

\begin{figure}
    \centering
    \includegraphics[width=0.48\textwidth]{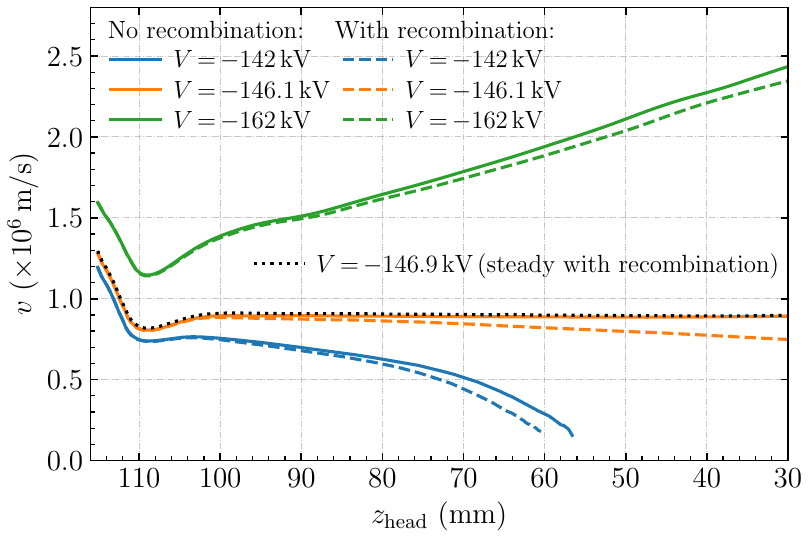}
    \caption{Comparison of negative streamer propagation with and without the additional reactions given in~\tref{tab:recombination}. The streamer velocity $v$ is shown versus the streamer position $z_{\mathrm{head}}$ for three different streamers corresponding to \fref{fig:three examples evolution}.
     With electron-ion recombination, a steady negative streamer is found at $V=-$146.9\,kV.}
    \label{fig:with and without recombination}
\end{figure}

We here investigate the effect of including the additional positive ion conversion and electron-ion recombination reactions given in~\tref{tab:recombination}.
\Fref{fig:with and without recombination} shows how accelerating, steady and negative streamers are affected by including these reactions.
The simulations were performed in domain A, see section~\ref{sec:three examples}.
With the extra reactions, negative streamers are slightly slower due to increased conductivity loss behind the streamer head.
However, the effect is rather small, and the steady applied voltage changes by only 0.5\%.

\begin{table*}
\centering
\captionsetup{width=0.92\textwidth}
\caption{\label{tab:recombination}Additional positive ion conversion and electron-ion recombination reactions.
$T$(K) and $T_e$(K) are gas and electron temperatures, respectively.
$T_e$ is computed as $T_e = 2\epsilon_\mathrm{e} / 3k_\mathrm{B}$ with the mean electron energy $\epsilon_\mathrm{e}$ obtained from BOLSIG+~\cite{hagelaar2005}.}
\begin{tabular*}{0.92\textwidth}{c@{\extracolsep{\fill}}llc}
\br
No. & Reaction & Reaction rate coefficient & Reference\\
\mr
\multicolumn{4}{l}{Positive ion conversion}\\
R14 & $\mathrm N_2^+ + \mathrm O_2 \to \mathrm O_2^+ + \mathrm N_2$ & $k_{14}=6.0\times10^{-17}(\frac{300}{T})^{0.5}\,\mathrm{m^3\,s^{-1}}$ & \cite{kossyi1992}\\
R15 & $\mathrm N_2^+ + \mathrm N_2 + \mathrm M \to \mathrm N_4^+ + \mathrm M$ & $k_{15}=5.0\times10^{-41}(\frac{300}{T})^{2.2}\,\mathrm{m^6\,s^{-1}}$ & \cite{kossyi1992} \\
R16 & $\mathrm N_4^+ + \mathrm O_2 \to \mathrm O_2^+ + \mathrm N_2 + \mathrm N_2$ & $k_{16}=2.5\times10^{-16}\,\mathrm{m^3\,s^{-1}}$ & \cite{kossyi1992} \\
R17 & $\mathrm O_2^+ + \mathrm O_2 + \mathrm M \to \mathrm O_4^+ + \mathrm M$ & $k_{17}=2.4\times10^{-42}(\frac{300}{T})^{3.2}\,\mathrm{m^6\,s^{-1}}$ & \cite{kossyi1992} \\
\mr
\multicolumn{4}{l}{Electron-ion recombination}\\
R18 & $\mathrm e + \mathrm N_2^+ + \mathrm M \to \mathrm N_2 + \mathrm M$ & $k_{18}=6.0\times10^{-39}(\frac{300}{T_e})^{1.5}\,\mathrm{m^6\,s^{-1}}$ & \cite{kossyi1992}\\
R19 & $\mathrm e + \mathrm O_2^+ + \mathrm M \to \mathrm O_2 + \mathrm M$ & $k_{19}=6.0\times10^{-39}(\frac{300}{T_e})^{1.5}\,\mathrm{m^6\,s^{-1}}$ & \cite{kossyi1992} \\
R20 & $\mathrm e + \mathrm N_2^+ \to \mathrm N_2$ & $k_{20}=1.5\times10^{-13}\,\mathrm{m^3\,s^{-1}}$ & \cite{tochikubo2002} \\
R21 & $\mathrm e + \mathrm O_2^+ \to \mathrm O_2$ & $k_{21}=1.5\times10^{-13}\,\mathrm{m^3\,s^{-1}}$ & \cite{tochikubo2002} \\
R22 & $\mathrm e + \mathrm N_4^+ \to \mathrm N_2 + \mathrm N_2$ & $k_{22}=2.0\times10^{-12}(\frac{300}{T_e})^{0.5}\,\mathrm{m^3\,s^{-1}}$ & \cite{kossyi1992} \\
R23 & $\mathrm e + \mathrm O_4^+ \to \mathrm O_2 + \mathrm O_2$ & $k_{23}=1.4\times10^{-12}(\frac{300}{T_e})^{0.5}\,\mathrm{m^3\,s^{-1}}$ & \cite{kossyi1992} \\
\br
\end{tabular*}
\end{table*}

\begin{figure}
    \centering
    \includegraphics[width=0.48\textwidth]{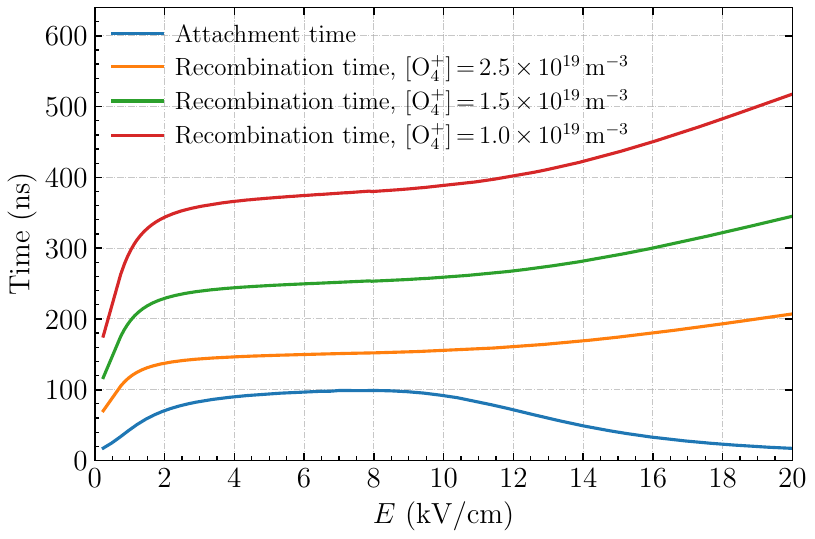}
    \caption{The electron attachment time and $e + \mathrm O_4^+$ recombination time versus the electric field $E$.
    The $e + \mathrm O_4^+$ recombination time depends on the $\mathrm O_4^+$ density, which in our simulations was generally below $10^{19} \, \mathrm{m}^{-3}$.}
    \label{fig:attachment and recombination times}
\end{figure}

O$_4^+$ is typically the dominant positive ion in streamer discharges at atmospheric pressure air.
In \fref{fig:attachment and recombination times}, the electron attachment time is compared with the $e + \mathrm O_4^+$ recombination time, where these time scales are given by the inverse of the respective reaction frequencies.
% The attachment time is the inverse of the attachment frequency, and it is below 100\,ns.
% The $e + \mathrm O_4^+$ recombination time is the inverse of the $e + \mathrm O_4^+$ recombination frequency, which depends on the electric field and the $\mathrm O_4^+$ density.
The $e + \mathrm O_4^+$ recombination time depends on the $\mathrm O_4^+$ density, and in our simulations, $\mathrm O_4^+$ densities were generally below $10^{19} \, \mathrm{m}^{-3}$. % so $e + \mathrm O_4^+$ recombination times are several hundred ns.
Including positive ion conversion and electron-ion recombination -- which we in hindsight maybe should have done -- would therefore not significantly alter the results presented in this paper.

In contrast, for positive streamers in air, the $e + \mathrm O_4^+$ recombination time can be shorter than the electron attachment time due to a low internal electric field and high $\mathrm O_4^+$ density, so that electron-ion recombination plays a significant role~\cite{niknezhad2021b, starikovskiy2021, francisco2021a, francisco2021e, li2022}.

\section*{References}
\normalem
\bibliography{references}

\end{document}